\documentclass[12pt]{article}

\pdfoutput=1
\usepackage{graphicx}
\usepackage{amssymb}
\usepackage{amsmath}
\usepackage{color}
\usepackage{subfigure}
\usepackage[colorlinks=true,linkcolor=blue,citecolor=blue]{hyperref}

\begin{document}

\newcommand{\be}{\begin{equation}}
\newcommand{\ee}{\end{equation}}


\begin{titlepage}

\begin{flushright}
ICRR-Report-639-2012-28  \\
IPMU 12-6216\\
UT-12-36
\end{flushright}

\vskip 2cm

\begin{center}

{\Large \sf
Smooth hybrid inflation in a supersymmetric axion model
}

\vspace{1cm}

{Masahiro Kawasaki$^{(a, b)}$, Naoya Kitajima$^{(a)}$
and Kazunori Nakayama$^{(c,b)}$}

\vskip 1.0cm

{\it
$^a$Institute for Cosmic Ray Research,
     University of Tokyo, Kashiwa, Chiba 277-8582, Japan\\
$^b$Kavli Institute for the Physics and Mathematics of the Universe, 
     University of Tokyo, Kashiwa, Chiba 277-8568, Japan\\
$^c$Department of Physics, University of Tokyo, Bunkyo-ku, Tokyo 113-0033, Japan
}

\vskip 1.0cm

\begin{abstract}
	We show that the smooth hybrid inflation is naturally realized in a framework 
	of supersymmetric axion model.
	Identifying the Peccei-Quinn scalar fields as a part of the infaton sector, 
	successful inflation takes place reproducing the amplitude and spectral index 
	of the curvature perturbation observed by WMAP.
	A relatively large axion isocurvature perturbation and its non-Gaussianity are 
	predicted in our model.
	The saxion coherent oscillation has a large amplitude and dominates 
	the Universe. 
	The subsequent decay of the saxion produces huge amount of entropy, which dilutes
	unwanted relics. 
	Winos, the lightest supersymmetric particles in this scenario, 
	are produced non-thermally  
	in the decay and account for dark matter.
\end{abstract}

\end{center}

\end{titlepage}

\newpage
\tableofcontents

\vspace{1cm}

\section{Introduction} \label{intro}

The Standard Model (SM) in particle physics is confirmed after the recent discovery 
of the Higgs boson at the LHC~\cite{:2012gk}.
However, the SM is not believed to be a fundamental theory because of some 
theoretical problems.
One of these is known as the strong CP problem.
Although quantum chromodynamics (QCD) allows the existence of the CP-violating 
term in the Lagrangian, the experimental test of the neutron electric dipole moment 
suggests that CP must be preserved with very high accuracy implying that the 
CP-violating term must be extremely small~\cite{Baker:2006ts}.
The only plausible solution to the the strong CP problem so far is the Peccei-Quinn 
(PQ) mechanism~\cite{Peccei:1977hh}.
In the PQ mechanism, a global symmetry written as $U(1)_\mathrm{PQ}$
is spontaneously broken and a Pseudo-Nambu-Goldstone boson arises accordingly : 
it is called axion.
The axion is assumed to acquire the potential dominantly through the QCD instanton 
effect, which drives the axion to the CP-preserving minimum~\cite{Kim:1986ax}.

On the other hand, the recent cosmological observations,
including the Wilkinson Microwave Anisotropy Probe (WMAP) 
observations~\cite{Komatsu:2010fb} 
strongly support inflation in the very early Universe.
Generally, inflation is considered to be driven by some scalar field called inflaton with almost flat potential.
Such a scalar field is not predicted in the framework of SM.
It is also revealed that our present Universe is filled with the non-baryonic 
cold dark matter (CDM), whose density parameter is found to be~\cite{Komatsu:2010fb}
\be
	\Omega_\mathrm{CDM} h^2 = 0.1126 \pm 0.0036,
\ee
where $h$ is the dimensionless Hubble parameter in units of 
$H_0 = 100 h ~\mathrm{km/sec/Mpc}$ with $H_0$ being the present Hubble parameter.
The SM does not contain any candidate of the CDM, and hence we are forced 
to go beyond the SM.
Supersymmetric(SUSY) theories generally predict many scalar fields with flat potentials 
one of which may be an inflaton, and SUSY protects flatness of the potential 
against large radiative corrections. 
Furthermore, the lightest SUSY particle is a good candidate for dark matter 
due to R-parity conservation.  

Motivated by these observations, we consider a SUSY axion model
in order to account for inflation and DM with solving the strong CP problem in a unified framework.
In the previous papers~\cite{Kawasaki:2010gv,Kawasaki:2011ym}, 
we studied the inflation and the following cosmological history in a SUSY axion model.
We pointed out that the SUSY hybrid inflation~\cite{Copeland:1994vg} naturally takes place by identifying the PQ scalar fields as waterfall fields.
To reproduce the observed curvature perturbation, the PQ scale $(f_a)$ must be of order of $f_a\sim 10^{15}\,\mathrm{GeV}$.
Generally, such a large PQ scale leads to the overproduction of axions because the amplitude of the axion coherent oscillation
scales as the PQ scale.
However, considering the post-inflationary dynamics, the saxion (the scalar partner of the axion) dominates the Universe 
and the subsequent decay of the saxion produces a huge amount of entropy.
The axion density is diluted and, as a result, it can have a right dark matter
abundance~\cite{Steinhardt:1983ia,Kawasaki:1995vt}.
The baryon asymmetry is also diluted by the entropy production.
However, adopting the Affleck-Dine mechanism \cite{Affleck:1984fy} for baryogenesis, a large initial baryon asymmetry enough to survive the dilution can be generated.

In this paper, we consider a variant inflation model: smooth hybrid inflation model~\cite{Lazarides:1995vr,Yamaguchi:2004tn} 
in the framework of the SUSY axion model by modifying the PQ sector appropriately.
The idea was already noted shortly in the previous paper~\cite{Kawasaki:2011ym}, but complete analyses were not yet performed.
We will see that the successful smooth hybrid inflation takes place in a SUSY axion model.
Furthermore, considering the post-inflationary dynamics, we show that the saxion coherent oscillation inevitably dominates the Universe and the subsequent decay of the saxion can produce a correct amount of dark matter.

This paper is organized as follows.
In Sec.~\ref{SUSY_axion}, we introduce a SUSY axion model we have in mind.
In Sec.~\ref{SHI}, the smooth hybrid inflation model is reviewed.
In Sec.~\ref{dynamics}, reheating after inflation and the dynamics of saxion and its cosmological implications are discussed.
In Sec.~\ref{isocurvature}, we show the constraint from the isocurvature perturbation and calculate the non-Gaussianity.
We conclude in Sec.~\ref{conc}.

\section{A supersymmetric axion model} \label{SUSY_axion}

\subsection{The potential of the SUSY axion model} \label{potential}

In this section, we briefly review the SUSY axion model.
We consider the SUSY axion model whose superpotential is given by
\be
	W = S \bigg( -\mu^2 + \frac{(\Psi \bar\Psi)^n}{M^{2(n-1)}} \bigg)  + \lambda \Psi X \bar X,
	\label{superpot_axion}
\ee
where $S$ is a gauge-singlet chiral superfield and $\Psi$ and $\bar\Psi$ are the chiral PQ superfields which are also gauge-singlet.
$X$ and $\bar X$ are the chiral superfields interacting with the PQ field at tree level and they also interact with the minimal supersymmetric standard model (MSSM) fields through the gauge interaction.
$\mu$ and $M$ are some mass scales, $\lambda$ is a dimensionless coupling constant and $n$ is an integer larger than or equal to 2.
This superpotential has the global $U(1)_\mathrm{PQ}$ symmetry and the global $U(1)_R$ symmetry 
as well as the discrete $Z_n$ symmetry.
Charge assignments of respective fields are summarized in table~\ref{charge_assignments}.
The PQ superfields contain the axion ($a$), saxion ($\sigma$, the scalar partner of the axion) and axino ($\tilde a$, the fermionic superpartner of the axion).
There are two representative axion models : 
one is the Kim-Shifman-Vainshtein-Zakharov (KSVZ) model (also known as the hadronic axion model)~\cite{Kim:1979if} in which $X$ and $\bar X$ are additional heavy quarks, denoted by $Q$ and $\bar Q$.
The other is the Dine-Fischler-Srednicki-Zhinitsky (DFSZ) model~\cite{Dine:1981rt} in which $X$ and $\bar X$ are identified as
Higgs fields, $H_u$ and $H_d$.

\begin{table}[tb]
\begin{center}
	\begin{tabular}{|c|c|c|c|c|c|} \hline
		\rule[0mm]{0mm}{4.5mm} & $S$ & $\Psi$ & $\bar\Psi$ & $X$ & $\bar X$ \\ \hline \hline
		$U(1)_{\mathrm{PQ}}$ & 0 & $+1$ & $-1$ & $-1/2$ & $-1/2$ \\ \hline
		$U(1)_R$ & +2 & 0 & 0 & +1 & +1 \\ \hline
		$Z_n$       &  0 & 0 & $+1$ & 0 & 0 \\ \hline
	\end{tabular}
\end{center}
\caption{Charge assignments on the field content.}
\label{charge_assignments}
\end{table}

In the global SUSY limit, the $F$-term scalar potential is calculated as 
\be
	V_F =  \bigg| - \mu^2 + \frac{ (\Psi \bar\Psi)^n }{ M^{2(n-1)} } \bigg|^2 + \frac{ n^2 |S|^2 | \Psi \bar\Psi |^{2(n-1)} }{ M^{4(n-1)} } ( |\Psi|^2 + | \bar\Psi |^2 ),
\ee
where the scalar fields are denoted by same symbols as corresponding superfields.
Here and hereafter, $X$ and $\bar X$ are set to be zero due to Hubble-induced masses or thermal corrections
because $X=\bar X=0$ is en enhanced symmetry point.
The global minimum of the potential is placed at 
\be
	S=0 ~~~\text{and} ~~~\Psi \bar\Psi = f_a^2,
	\label{global_min}
\ee
where the PQ breaking scale $f_a$ is given by 
\be
	f_a = \big( \mu M^{n-1} \big)^{1/n}.
\ee
This indicates the existence of the flat direction along which the PQ fields do not feel the potential, ensured by the fact that 
the $U(1)_\mathrm{PQ}$ symmetry is extended to the complex $U(1)$ due to the holomorphy of the superpotential~\cite{Kugo:1983ma}.

The flat direction is lifted by the low-energy SUSY-breaking mass terms
\be
	V_\mathrm{soft} = c_1 m_{3/2}^2 |\Psi|^2 + c_2 m_{3/2}^2 |\bar\Psi|^2,
\ee
where $m_{3/2}$ is the gravitino mass and $c_1$ and $c_2$ are real, positive and $O(1)$ numerical constants.
This stabilizes the radial component of the PQ fields, $|\Psi|$ and $|\bar\Psi|$ at 
\be
	v \simeq \bigg( \frac{c_2}{c_1} \bigg)^{1/4} f_a,~~~ \bar v \simeq \bigg( \frac{c_1}{c_2} \bigg)^{1/4} f_a
	\label{minimum}
\ee
respectively.
The saxion is defined as the deviation from the minimum along the flat direction.
Near the minimum of the potential (\ref{minimum}), the axion and the saxon is related to the PQ fields as 
\be
	\Psi = v \exp \bigg( \frac{\sigma + ia}{\sqrt 2 F_a} \bigg) ,~~~ \bar\Psi = \bar v \exp \bigg( -\frac{\sigma + ia}{\sqrt 2 F_a} \bigg),
\ee
where $F_a$ is defined as $F_a = \sqrt{ v^2 + \bar v^2}$.
The PQ fields obtain vacuum expectation values (VEVs) and, since the $U(1)_{\rm PQ}$ is anomalous under the QCD,
the axion obtains an instanton-induced potential and solves the strong CP problem via the PQ mechanism.

\subsection{The decay of the saxion} \label{decay}

Because the decay of the saxion is an important ingrediate in the following discussion, we here summarize the decay rate of the saxion.
The saxion has an interaction with the axion through the kinetic terms of the PQ scalar fields as 
\be
	|\partial_\mu \Psi|^2 + |\partial_\mu \bar\Psi|^2 = \bigg( 1 + \frac{\sqrt 2 \xi}{F_a} \sigma \bigg) \bigg( \frac{1}{2} (\partial_\mu a)^2 + \frac{1}{2} (\partial_\mu\sigma)^2 \bigg) + \dots,
\ee
where $\xi$ is defined as $\xi \equiv (v^2 - \bar v^2 )/F_a^2$ which is generally of order unity unless $v \simeq \bar v$ i.e. $c_1 \simeq c_2$ \cite{Chun:1995hc,Kawasaki:2007mk}.
The decay rate of the saxion into the axion pair is derived as
\be
	\Gamma_{\sigma \to aa} = \frac{\xi^2}{64 \pi} \frac{m_\sigma^3}{F_a^2},
\ee
where $m_\sigma$ is the mass of the saxion which is same order of the gravitino mass.

In the KSVZ axion model with only one pair of fundamental and anti-fundamental representation of SU(3) ($Q$ and $\bar Q$), 
the main decay mode of the saxion into the MSSM particles is that into two gluons whose decay rate is calculated as
\be
	\Gamma_{\sigma \to gg} = \frac{\alpha_s^2}{64 \pi^3} \frac{m_\sigma^3}{F_a^2},
	\label{sigma-gg}
\ee
where $\alpha_s$ is the QCD gauge coupling constant.\footnote{
	The axion decay constant, $f_{\rm PQ}$, is given by $f_{\rm PQ}=\sqrt{2}F_a/N_{\rm DW}$ with $N_{\rm DW}$
	being the domain wall number, which depends on the model.
	Taking the model-dependence from $N_{\rm DW}$, $c_1$ and $c_2$ into account, we do not distinguish 
	$f_{\rm PQ}$ and $f_a$ hereafter.
}
Another efficient decay mode of the saxion is that into two gluinos if it is allowed kinematically, whose decay rate is given by
\be
	\Gamma_{\sigma \to \tilde g \tilde g} = \frac{\alpha_s^2}{64 \pi^3} |d|^2 \frac{m_\sigma^3}{F_a^2},
	\label{sigma-gluino}
\ee
where $d$ is an $O(1)$ numerical constant for $m_\sigma \sim m_{3/2}$ \cite{Endo:2006ix}.
In the KSVZ axion model, the decay mode into the MSSM particles may be subdominant and the dominant decay mode may be the two axion decay unless $c_1 \simeq c_2$.
Assuming the decay into two axion is suppressed, the decay temperature is calculated as
\be
	T_\sigma \simeq 100\,\mathrm{MeV} \bigg(\frac{m_\sigma}{100\,\mathrm{TeV}} \bigg)^{3/2} \bigg( \frac{10^{15}\,\mathrm{GeV}}{F_a} \bigg).
	\label{T_sigma_K}
\ee

In the DFSZ axion model, there exists the tree level coupling of the saxion with the standard model Higgses, so the saxion decays dominantly into the Higgses.
The decay rate of the saxion into the lightest Higgs boson pair is derived as
\be
	\Gamma_{\sigma \to hh} = \frac{1}{16 \pi} \frac{m_\sigma^3}{F_a^2} \bigg( \frac{\mu}{m_\sigma} \bigg)^4 
	\label{sigma-hh}
\ee
where $\mu$ is the higgsino mass defined via $\mu = \lambda v$\footnote{
	In our model, $\lambda$ must be an extremely small value such as $\lambda \sim 10^{-12}$ for $v \sim 10^{15}~\mathrm{GeV}$ in order to realize the appropriate $\mu$-term.
	To avoid such a small $\lambda$, we can adopt the superpotential $W = \lambda \Psi^{m}/M_P^{m-1}H_d H_d$ with some integer $m \geq 2$ instead of the second term in (\ref{superpot_axion}).
	Adopting such a superpotential, the decay rate of the saxion into the Higgs boson pair and that into the higgsino pair are $m^2$ times larger than (\ref{sigma-hh}) and (\ref{sigma-higgsino}).
}.
The decay of the saxion into higgsinos is also efficient if it is allowed kinematically and its decay rate is given by
\be
	\Gamma_{\sigma \to \tilde h \tilde h} = \frac{1}{8 \pi} \frac{m_\sigma^3}{F_a^2} \bigg( \frac{\mu}{m_\sigma} \bigg)^2.
	\label{sigma-higgsino}
\ee
In the DFSZ axion model, the decay mode into the MSSM particles may dominate over the two axion decay.
The saxion decay temperature is then calculated as
\be
	T_\sigma \simeq 5\,\mathrm{GeV} \bigg( \frac{m_\sigma}{100\,\mathrm{TeV}} \bigg)^{3/2} \bigg( \frac{10^{15}\,\mathrm{GeV}}{F_a} \bigg) \bigg( \frac{\mu}{m_\sigma} \bigg)^2.
	\label{T_sigma_D}
\ee
Note also that the mixing of the PQ scalar with Higgs fields induce the saxion decay into SM quarks and leptons.
They are subdominant as long as the saxion is heavy ($m_\sigma \gtrsim 1$\,TeV).

\section{Smooth hybrid inflation in a SUSY axion model} \label{SHI}

\subsection{The potential of the inflaton}

In this section, we review the smooth hybrid inflation model proposed originally in~\cite{Lazarides:1995vr},
including the supergravity correction.
We start with the PQ sector superpotential (\ref{superpot_axion}) and the following K\"ahler potential,
\be
	W = S \bigg( -\mu^2 + \frac{(\Psi \bar\Psi)^n}{M^{2(n-1)}} \bigg) + W_0,
	\label{superpot}
\ee
\be
	K = |S|^2 + |\Psi|^2 + |\bar\Psi|^2 + \dots,
	\label{Kahler_pot}
\ee
where $\dots$ in the K\"ahler potential denotes the higher order Planck suppressed terms.
We have added a constant term $W_0$ in the superpotential, which is required to cancel the positive vacuum energy from the SUSY breaking and make the cosmological constant zero in the present vacuum.
The gravitino mass is related to it as $W_0 = m_{3/2} M_P^2$.
We take $M_P = 1$ in this section.

\begin{figure}[t]
\centering
\includegraphics [width = 10cm, clip]{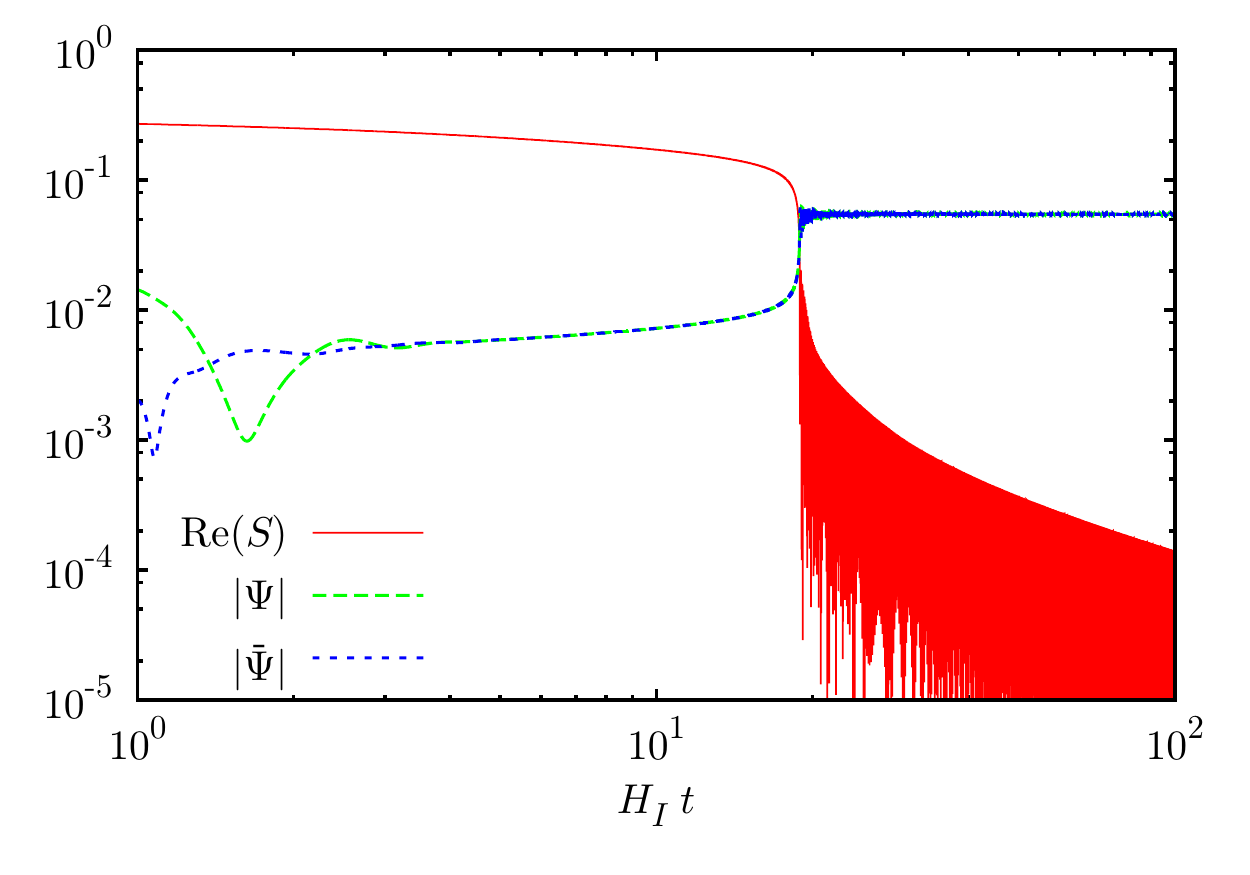}
\caption{
	The time evolution of $\mathrm{Re(S)}$ (red solid line), $|\Psi|$ (green dashed line) and $|\bar\Psi|$ (blue dotted line) in the inflationary epoch are shown.
	The x-axis is the time multiplied by the Hubble parameter during inflation.
	We have taken $\mu = 0.002$, $M = 1.5$ ($f_a = 0.05$), $n=2$ and $W_0 = 0$ and $S = 0.3$, $\Psi = 0.01+0.002i$, $\bar\Psi = 0.02 + 0.001i$ and $H=2 \times 10^{-6}$ as initial values in Planck units ($M_P = 1$). 
}
\label{Fig1}
\end{figure}

The $F$-term scalar potential is calculated by using the formula
\be 
	V_F = e^{K/M_P^2} \big[ K^{ij^*} D_i W  D_{j^*} W^* - 3 |W|^2 / M_P^2 \big],
\ee
where $D_i W = W_i + K_i W / M_P^2$ and the subscript represents the derivative with respect to corresponding field
and $K^{i j^*}$ is the inverse matrix of $K_{i j^*}$.
Then, the scalar potential is calculated as 
\be
	\begin{split}
		V = & e^{|S|^2 + |\Psi|^2 + |\bar\Psi|^2 } 
			\Biggl\{
				\biggl| (1+ |S|^2) \bigg(-\mu^2 + \frac{(\Psi \bar\Psi)^n}{M^{2(n-1)}}\bigg) + S^* W_0 \biggr|^2 \\[1mm]
				& + \biggl| nS \bar\Psi \frac{(\Psi \bar\Psi)^{n-1}}{M^{2(n-1)}}  + \Psi^* \biggl[ S \biggl( - \mu^2 + \frac{(\Psi \bar\Psi )^n}{M^{2(n-1)}} \biggr) + W_0 \biggr] \biggr|^2 \\[1mm]
				& + \biggl| nS \Psi \frac{(\Psi \bar\Psi)^{n-1}}{M^{2(n-1)}}  + \bar\Psi^* \biggl[ S \biggl( - \mu^2 + \frac{(\Psi \bar\Psi )^n}{M^{2(n-1)}} \biggr) + W_0 \biggr] \biggr|^2 \\[1mm]
				& - 3 \biggl| S \biggl( - \mu^2 + \frac{(\Psi \bar\Psi )^n}{M^{2(n-1)}} \biggr) + W_0 \biggr|^2 
			\Biggr\}.
	\end{split}
\ee
Writing down only a few significant terms, the scalar potential is expressed as
\be
	\begin{split}
		V = & \bigg( 1 + | \Psi |^2 + | \bar\Psi |^2 + \frac{1}{2} |S|^4 \bigg) \bigg| - \mu^2 + \frac{ (\Psi \bar\Psi)^n }{ M^{2(n-1)} } \bigg|^2 \\[1mm]
		& + \frac{ n^2 |S|^2 | \Psi \bar\Psi |^{2(n-1)} }{ M^{4(n-1)} } ( |\Psi|^2 + | \bar\Psi |^2 ) - 2 n \mu^2 |S|^2 \bigg( \frac{(\Psi \bar\Psi)^n}{ M^{2(n-1)} } + \mathrm{c.c.} \bigg) \\[1mm]
		& + 2W_0 \bigg[ S \bigg( \mu^2 + (n-1) \frac{(\Psi \bar\Psi)^n}{M^{2(n-1)}} \bigg) + \mathrm{c.c.} \bigg] + \dots.
		\label{scalar_pot2}
	\end{split}
\ee
If $|S| \gg (\mu M^{n-1})^{1/n} = f_a$ is hold, the PQ fields are placed on the following temporal minimum
\be
	|\Psi| = |\bar\Psi| \simeq \frac{f_a}{[n(2n-1)]^{1/2(n-1)}} \bigg( \frac{f_a}{|S|} \bigg)^{1/(n-1)} ~~~ \text{for} ~~~ |S| \gg f_a.
	\label{Psi_min_inf}
\ee
Expressing the complex inflaton field by $\varphi =\sqrt{2} |S|$ and $\theta_S = \mathrm{arg}(S)$ and substituting (\ref{Psi_min_inf}) into (\ref{scalar_pot2}), 
we get the effective potential of the inflaton $\varphi$ as
\be
	\begin{split}
		V (\varphi) =& \mu^4 \Bigg[ 1 -  \frac{4^{(2n-1)/(n-1)}}{[ 2n (2n-1) ]^{n/(n-1)}} \bigg( \frac{n-1}{4n-2} \bigg) \bigg( \frac{f_a}{\varphi} \bigg)^{2n/(n-1)} + \frac{1}{8} \varphi^4 +\dots \Bigg] \\[1mm]
		&+ 2 \sqrt{2} \mu^2 m_{3/2} \varphi \cos\theta_S \Bigg[ 1 + \frac{n-1}{[ n(n-1/2)]^{n/(n-1)}} \bigg( \frac{f_a}{\varphi} \bigg)^{2n/(n-1)} \Bigg]
		\label{inflaton_pot}
	\end{split}
\ee
for $|S| \gg f_a$.
Under this potential, the inflation takes place along the temporal valley given by (\ref{Psi_min_inf}) for large field value of $|S|$.
We numerically calculated the time evolutions of the fields $\mathrm{Re}(S)$, $|\Psi|$ and $|\bar\Psi|$ during and soon after the inflationary era, which are shown in Fig.~\ref{Fig1}.
This supports the above analytical calculations.

\begin{figure}[tp]
\centering
\subfigure[$n = 2$]{
	\includegraphics [width = 7.5cm, clip]{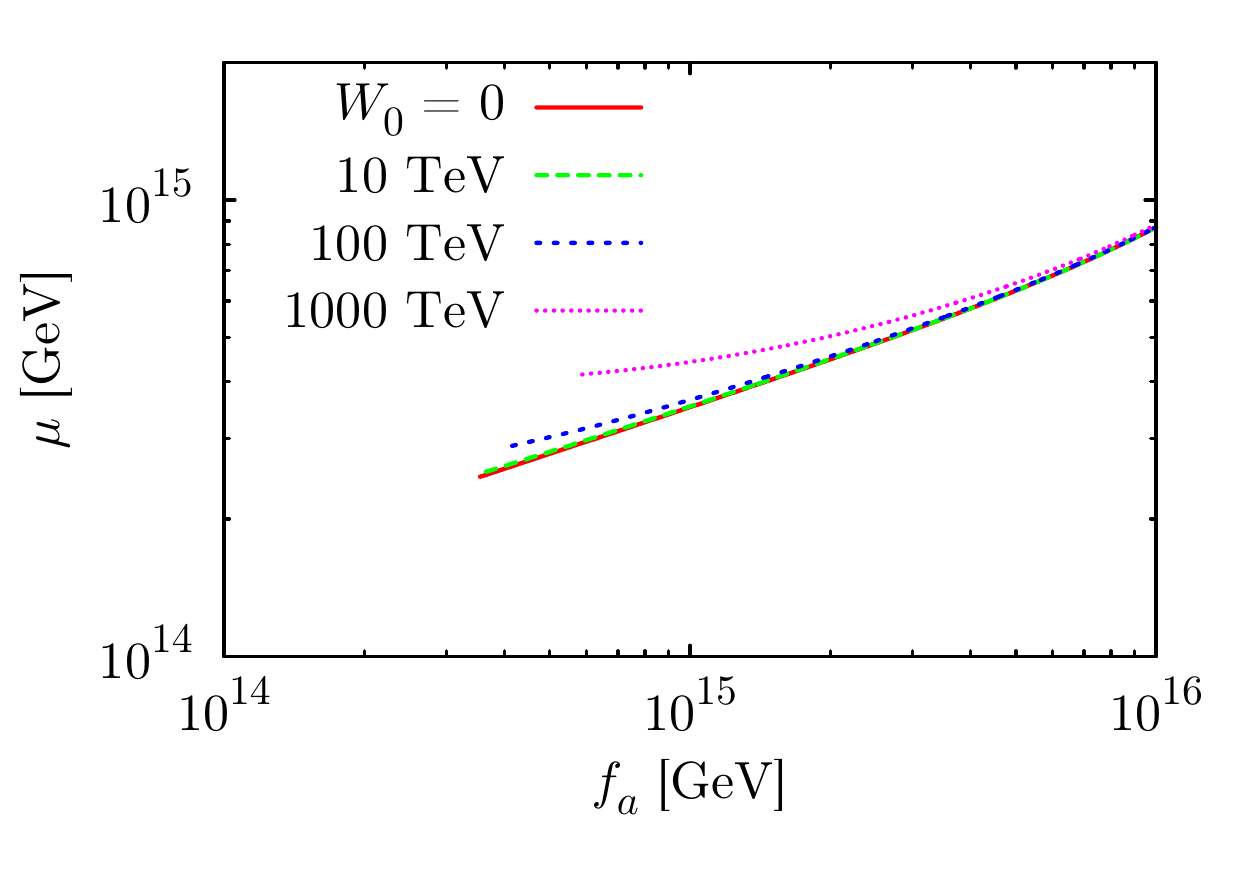}
	\label{Fig2c}
}
\subfigure[$n=2$]{
	\includegraphics [width = 7.5cm, clip]{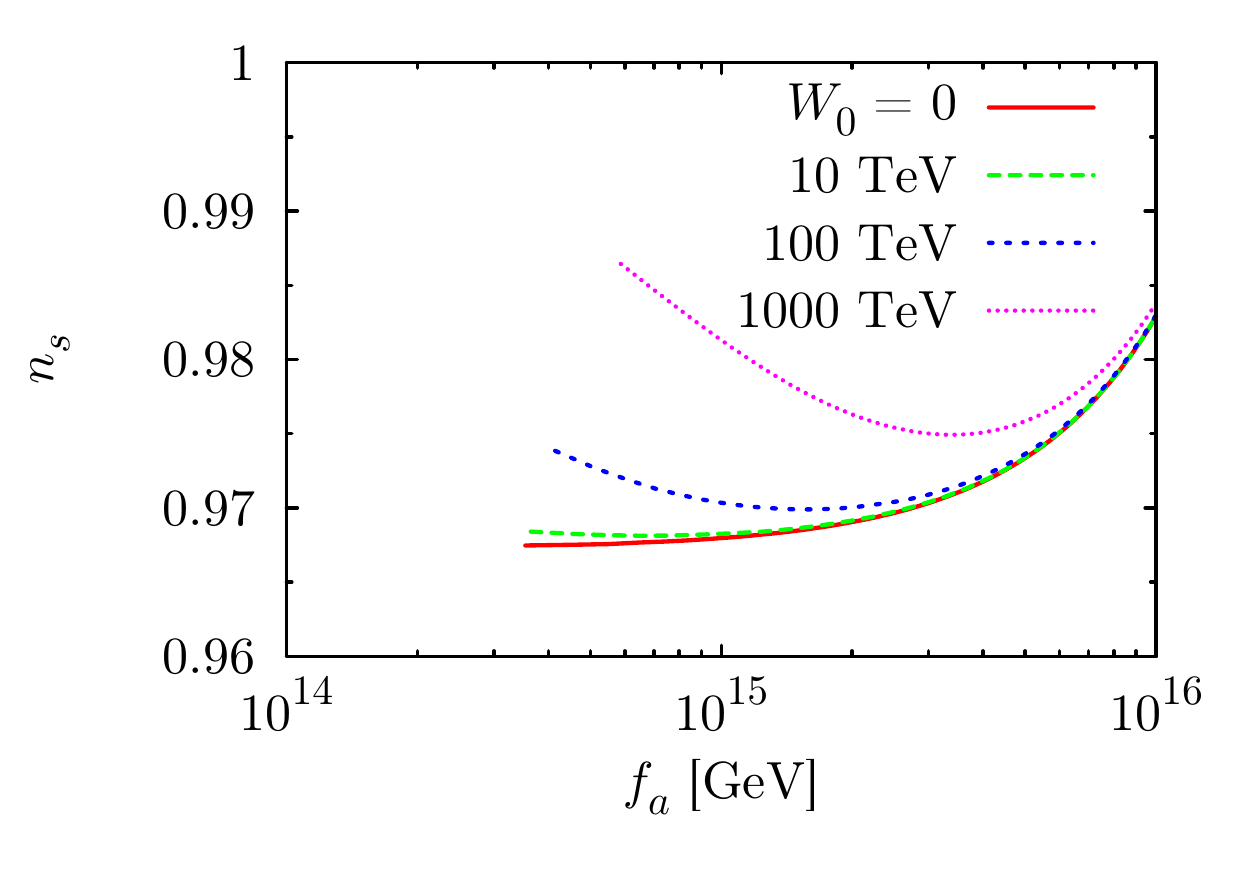}
	\label{Fig2d}
}
\subfigure[$n=4$]{
	\includegraphics [width = 7.5cm, clip]{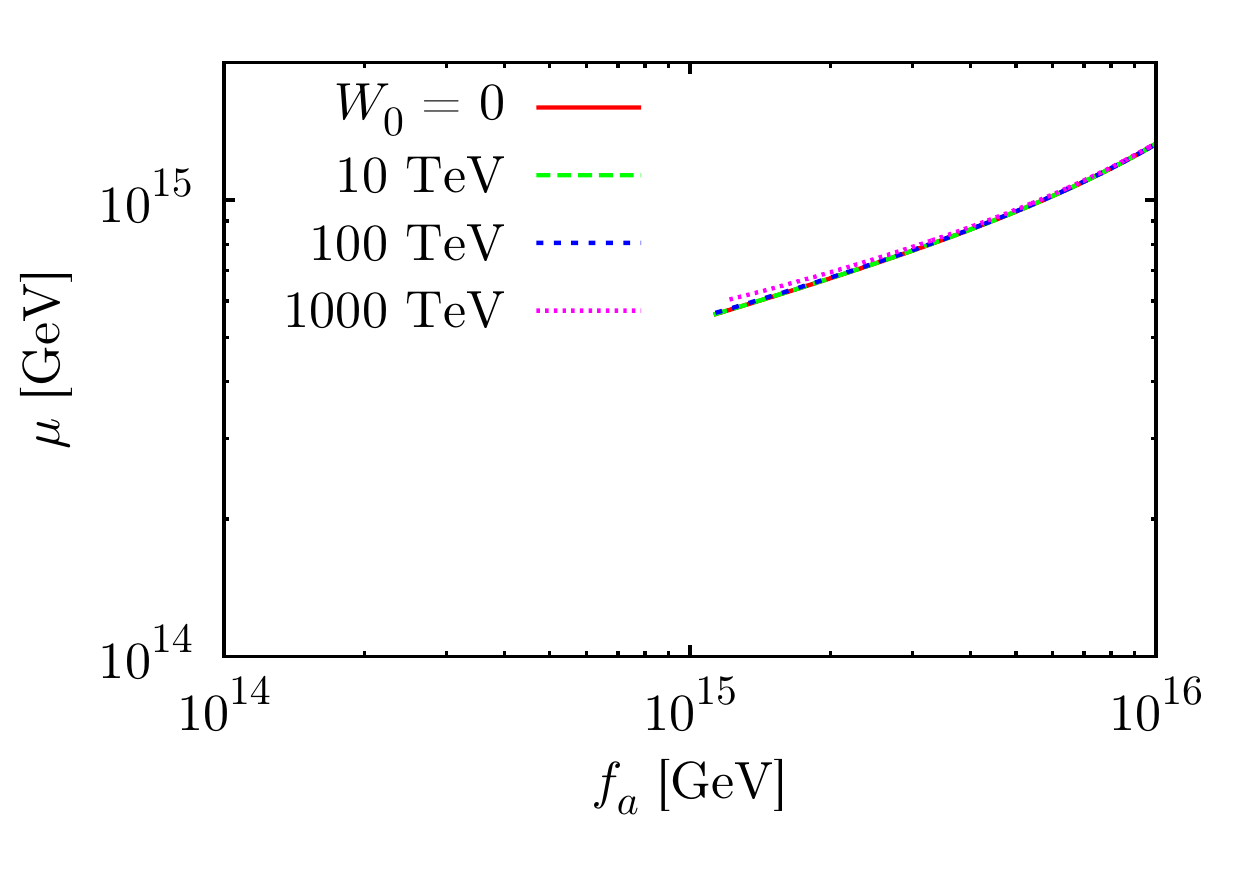}
	\label{Fig2e}
}
\subfigure[$n=4$]{
	\includegraphics [width = 7.5cm, clip]{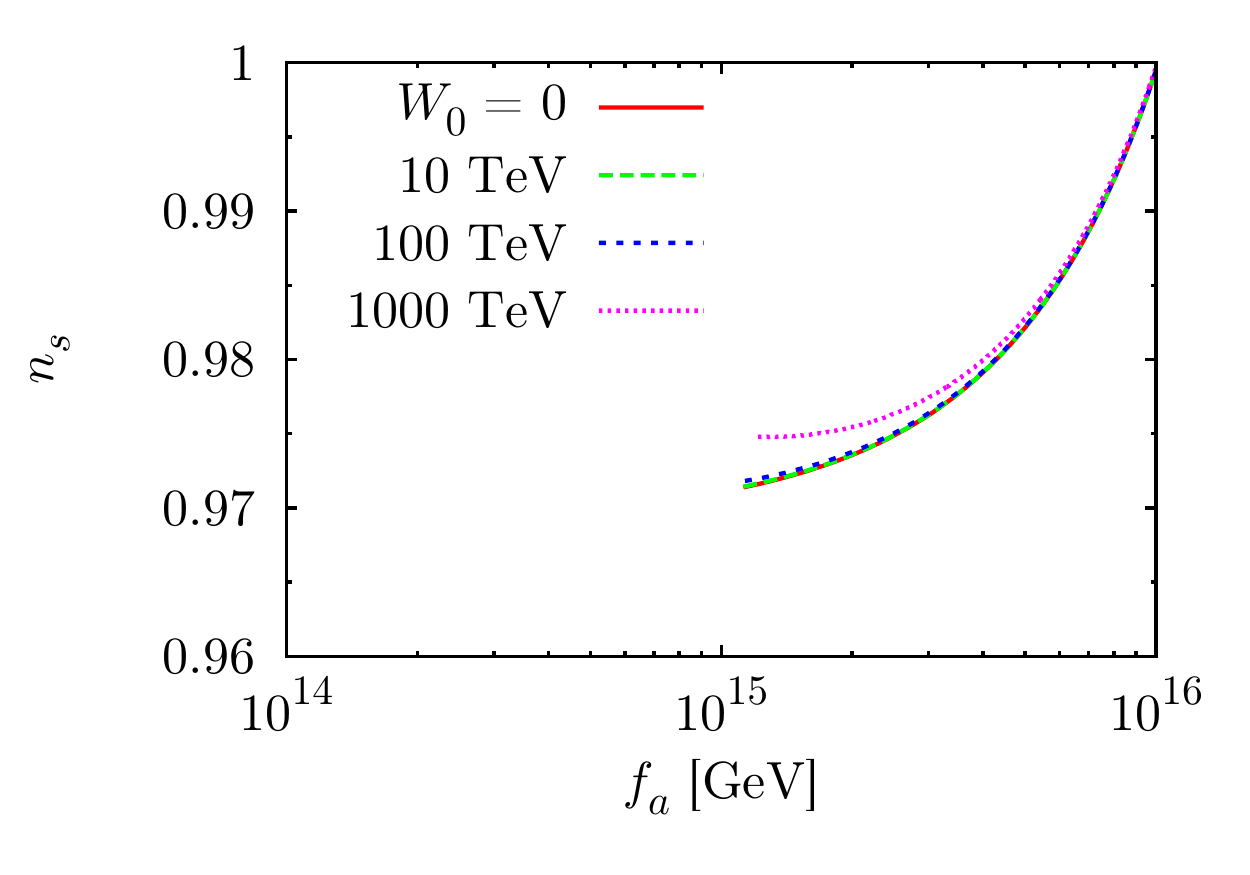}
	\label{Fig2f}
}
\caption{
	$\mu$ and $n_s$ as a function of $f_a$ are shown. The WMAP normalization of the density perturbation is imposed.
	We have taken $n=2$ (Fig.~\ref{Fig2c} and \ref{Fig2d}), $n=4$ (Fig.~\ref{Fig2e} and \ref{Fig2f}), $\theta_S = 0$, $W_0 = 0$ (solid-red line), $m_{3/2} = 10\,\mathrm{TeV}$ (dashed-green line), $m_{3/2} = 100\,\mathrm{TeV}$ (dotted-blue line) and $m_{3/2} = 1000\,\mathrm{TeV}$ (small-dotted-magenta line).
	The e-folding number is set to be 50 in all figures.
}
\label{Fig2}
\end{figure}

\begin{figure}[t]
\centering
\subfigure[n=2]{
	\includegraphics [width = 7.5cm, clip]{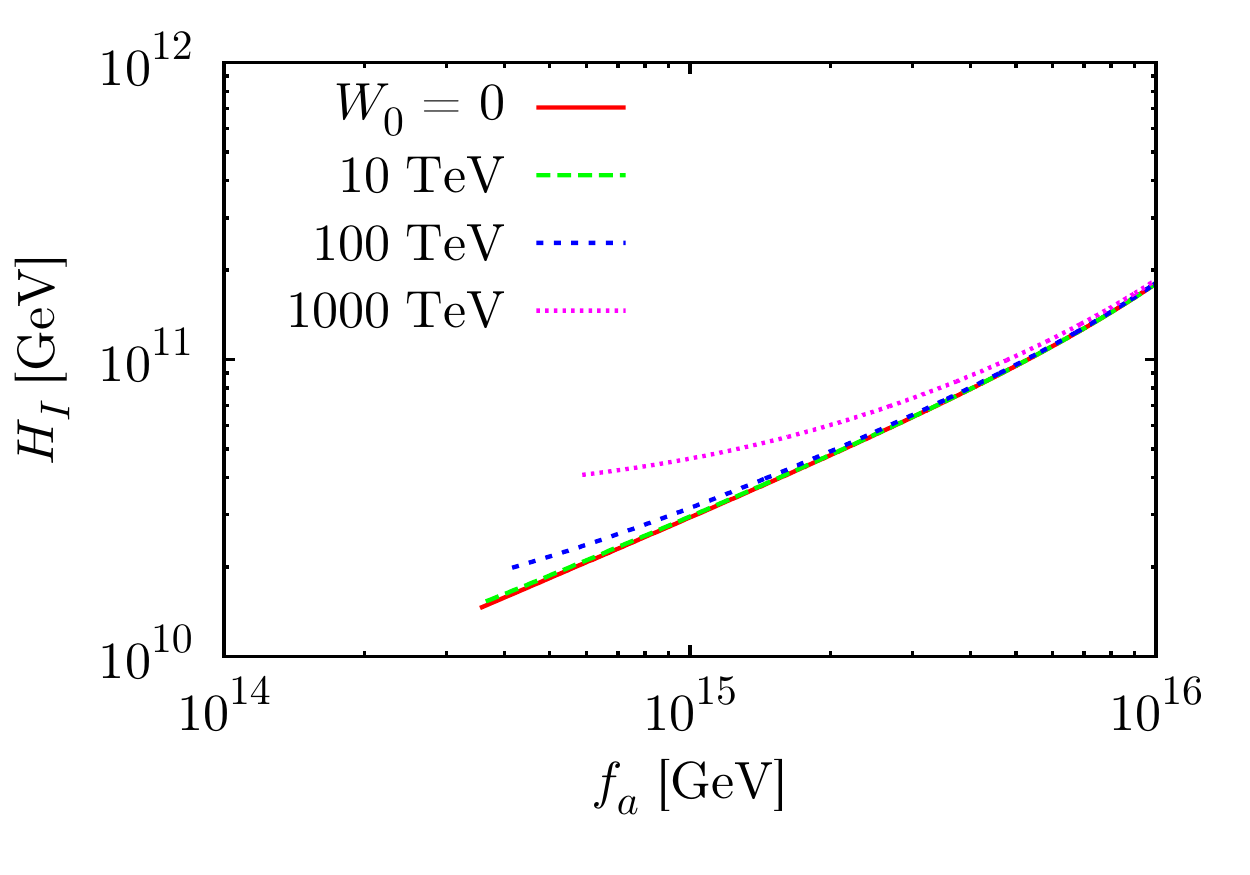}
	\label{Fig3a}
}
\subfigure[n=4]{
	\includegraphics [width = 7.5cm, clip]{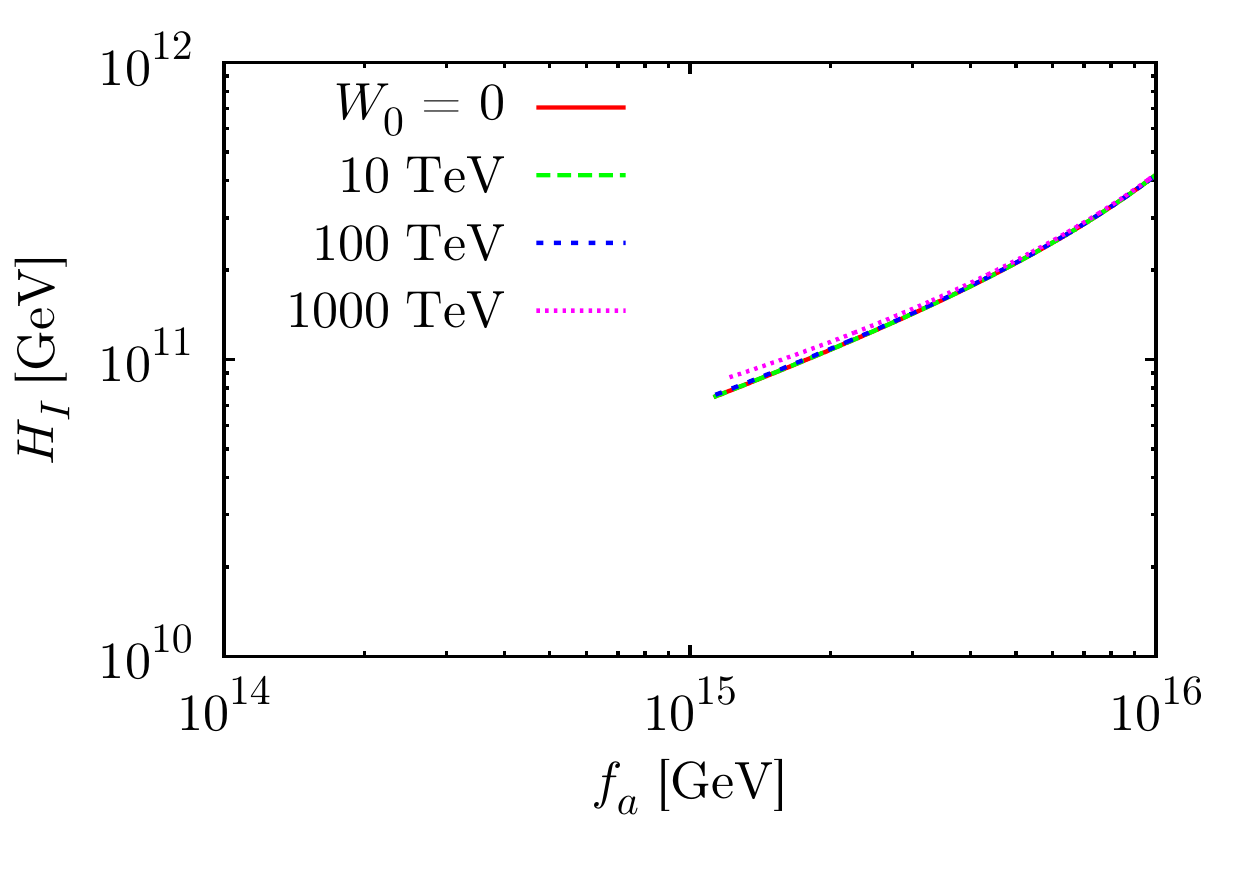}
	\label{Fig3b}
}
\caption{
	$H_I$ as a function of $f_a$ after the WMAP 
	normalization of the density perturbation is imposed.
	We have taken $n=2$ in Fig~\ref{Fig3a} and $n=4$ in Fig.~\ref{Fig3b} and $W_0=0$ (solid-red lines), $m_{3/2}=10~\mathrm{TeV}$ (dashed-green lines), $m_{3/2}=100~\mathrm{TeV}$ 
	(dotted-blue lines) and $m_{3/2}=1000~\mathrm{TeV}$ (small-dotted-magenta lines) and
	the e-folding number is set to be 50 in both figures.
}
\label{Fig3}
\end{figure}

From the inflaton potential (\ref{inflaton_pot}), the slow-roll parameters \cite{Liddle&Lyth} are calculated as 
\be
	\begin{split}
	\epsilon \equiv \frac{1}{2} \bigg( \frac{V'}{V} \bigg)^2
	=& \frac{1}{8} \Bigg[ \varphi_d^3 \bigg( \frac{\varphi_d}{\varphi} \bigg)^{(3n-1)/(n-1)} + \varphi^3 \\[1mm]
	&+ \frac{4 \sqrt{2} m_{3/2}}{\mu^2} \cos\theta_S \Bigg[ 1 - \frac{(n+1)(2n-1)}{8n} \varphi_d^4 \bigg( \frac{\varphi_d}{\varphi} \bigg)^{2n/(n-1)} \Bigg] \Bigg]^2,
	\end{split}
\ee
\be
	\begin{split}
		\eta \equiv \frac{V''}{V} =& -\frac{1}{2} \bigg( \frac{3n-1}{n-1} \bigg) \varphi_d^2 \bigg( \frac{\varphi_d}{\varphi} \bigg)^{(4n-2)/(n-1)} + \frac{3}{2} \varphi^2 \\[1mm]
		&+ \frac{\sqrt{2} (n+1)(2n-1)}{2(n-1)} \frac{m_{3/2}}{\mu^2} \cos\theta_S \varphi_d^3 \bigg( \frac{\varphi}{\varphi_d} \bigg)^{(3n-1)/(n-1)},
	\end{split}
\ee
where the prime denotes the derivative with respect to $\varphi$ and $\varphi_d$ is defined as 
\be
	\varphi_d = \left(\frac{4^{2n-1}}{2n(2n-1)^{2n-1}}\right)^{1/(6n-4)} f_a^{n/(3n-2)}.
\ee
The slow-roll condition is broken when $|\eta| \simeq 1$ at which $\varphi$ reaches the value given by
\be
	\varphi_c = \bigg[ \frac{n(3n-1)}{(n-1)(2n-1)} \bigg]^{(n-1)/(4n-2)} \frac{2 f_a^{n/(2n-1)}}{[2n(2n-1)]^{n/(4n-2)}}.
\ee
Using these quantities, we can calculate the power spectrum of the curvature perturbation and scalar spectral index $n_s$.
Results are shown in Fig.~\ref{Fig2}.
On each contour, the WMAP normalization of the density perturbation~\cite{Komatsu:2010fb} is imposed.
From these figures, the PQ breaking scale can be reduced to $\sim 4 \times 10^{14}\,\mathrm{GeV}$ and the spectral index well-agrees with the observational value.
Fig.~\ref{Fig3} shows the Hubble scale during inflation, $H_I$, as a function of $f_a$ by imposing the WMAP normalization.

\subsection{The initial value of the inflaton}

In some parameter regions, the linear term in the potential of the inflaton (\ref{inflaton_pot}) may create a 
local minimum if $\cos\theta_S < 0$, which may spoil the success of the inflation.\footnote{
	For the case of hybrid inflation, the effect of linear term was studied in Refs.~\cite{Buchmuller:2000zm,Senoguz:2004vu,Nakayama:2010xf}.
}
In such a case, the initial values of $\varphi$ and $\theta_S$ are constrained so that the inflaton is not trapped at the local minimum. 
Let us investigate these issues.

\begin{figure}[tp]
\centering
\subfigure[$n=2$]{
	\includegraphics [width = 7.5cm, clip]{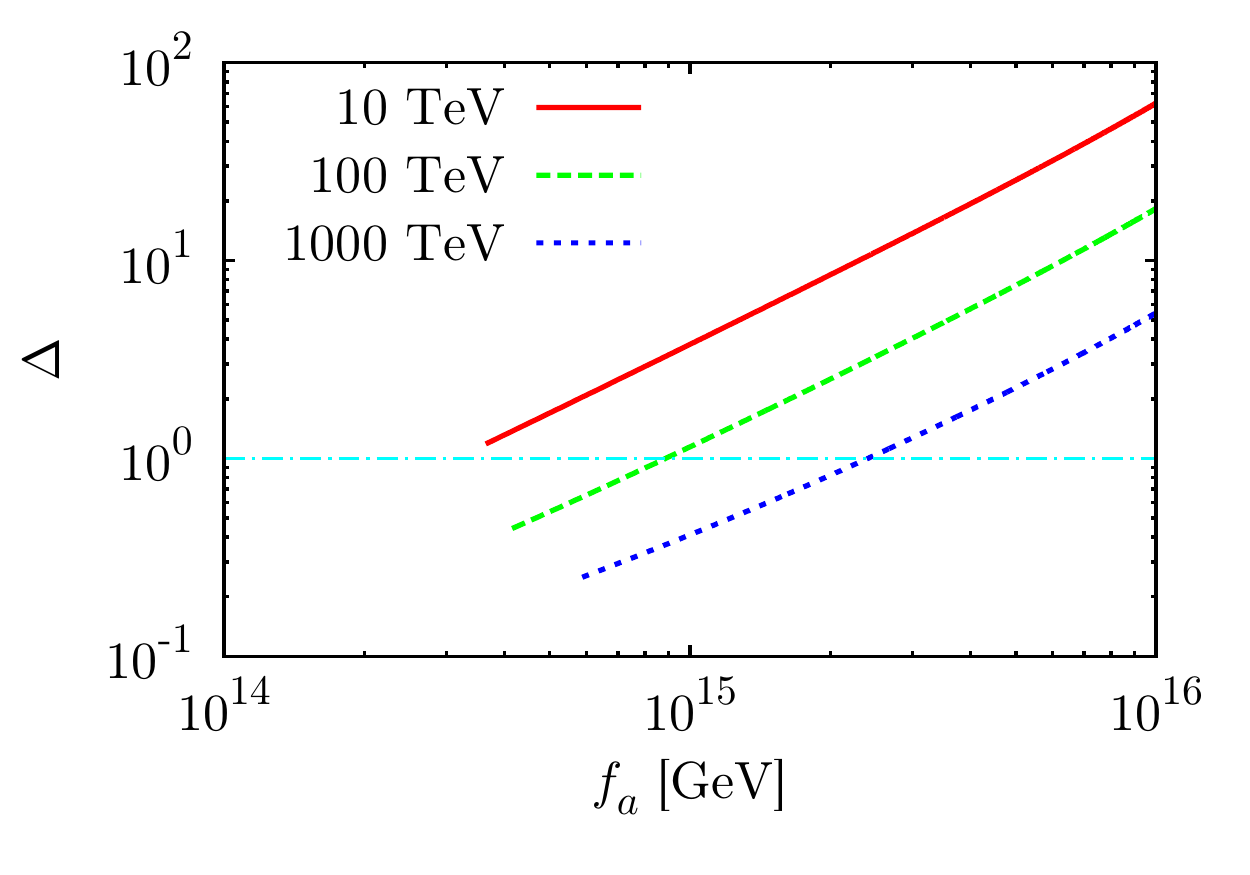}
	\label{Fig8a}
}
\subfigure[$n=4$]{
	\includegraphics [width = 7.5cm, clip]{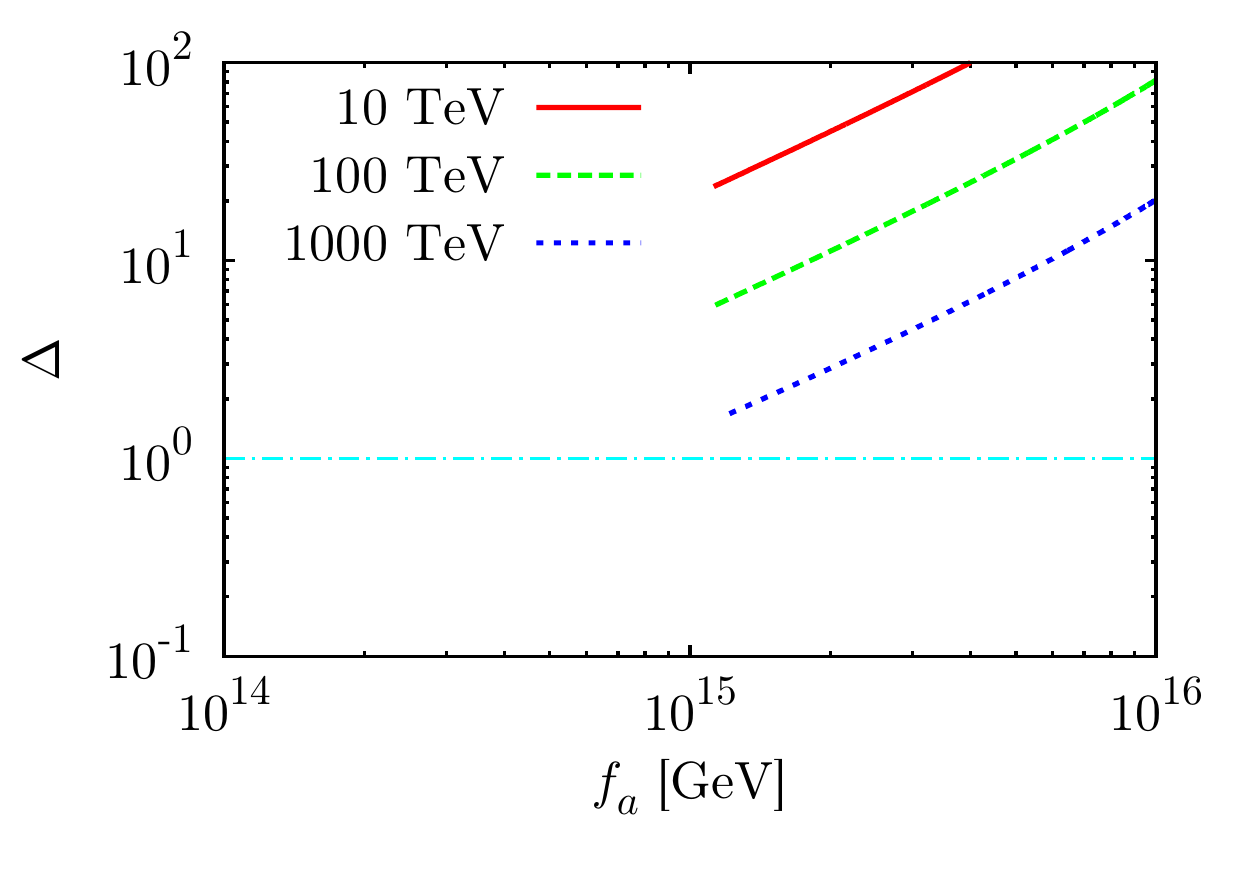}
	\label{Fig8b}
}
\caption{
	$\Delta$ parameters for $m_{3/2} = 10~\mathrm{TeV}$ (solid red lines), $100~\mathrm{TeV}$ (dashed green lines) and $1000~\mathrm{TeV}$ (dotted blue lines) are shown.
	We have taken $n=2$ (Fig.~\ref{Fig8a}) and $n=4$ (Fig.~\ref{Fig8b}) and imposed the CMB normalization condition.
}
\label{Fig8}
\end{figure}

First we examine whether a local minimum (and maximum) exists or not.
Clearly there is no such minimum for $\cos \theta_S \geq 0$ and hence we consider $\cos \theta_S < 0$ in the following.
For notational simplicity, let us rewrite the inflaton potential (\ref{inflaton_pot}) as 
\be
	V(\varphi) = V_0 - A \varphi^{-\alpha} - B \varphi + C \varphi^\beta,
	\label{inflaton_pot2}
\ee
where $V_0$, $A$, $B$, $C$, $\alpha$ and $\beta$ are defined as
\be
	\begin{split}
		&A = \frac{4^{(2n-1)/(n-1)}}{[2n(2n-1)]^{n/(n-1)}} \frac{n-1}{4n-2} \mu^4 f_a^\alpha,~~B = -2 \sqrt{2} \mu^2 m_{3/2} \cos\theta_S,~~C = \frac{\mu^4}{8} \\[2mm]
		&V_0 = \mu^4,~~\alpha = \frac{2n}{n-1},~~ \beta = 4.
	\end{split}
\ee
The derivative of the potential with respect to $\varphi$ is given by
\be
	V'(\varphi) = B \bigg[ \bigg(\frac{\varphi}{\varphi_\mathrm{max}} \bigg)^{-\alpha-1} + \bigg( \frac{\varphi}{\varphi_\mathrm{min}} \bigg)^{\beta-1} -1 \bigg],
\ee
where $\varphi_\mathrm{max}$ and $\varphi_\mathrm{min}$ are defined as
\be
	\varphi_\mathrm{max} = \bigg( \frac{\alpha A}{B} \bigg)^{1/(\alpha+1)},~~ \varphi_\mathrm{min} = \bigg( \frac{B}{\beta C} \bigg)^{1/(\beta-1)},
\ee
corresponding to the local maximum for small $C$ limit and the local minimum for small $A$ limit, respectively.
The condition for non-existing of the local maximum and the local minimum in the potential (\ref{inflaton_pot2}) is
\be
	V'(\varphi_V) > 0, 
	\label{no-exist_cond}
\ee
where $\varphi_V$ is defined through $V''(\varphi_V) = 0$ and given by
\be
	\varphi_V = \bigg( \frac{\alpha(\alpha+1) A}{\beta(\beta-1)C} \bigg)^{1/(\alpha + \beta)}.
\ee
Thus the condition (\ref{no-exist_cond}) is written as
\be
	\Delta \equiv \bigg( \frac{\alpha + \beta}{\beta-1}\bigg)^{\frac{\alpha+\beta}{(\alpha+1)(\beta-1)}} \bigg( \frac{\beta-1}{\alpha+1} \bigg)^{1/(\beta-1)} \frac{\varphi_\mathrm{max}}{\varphi_\mathrm{min}} > 1.
\ee
If this condition is satisfied, the local minimum of the potential does not appear and the initial value of $\theta_S$ is not constrained.
Moreover, there is no upper bound on the initial value of $\varphi$, and hence the initial value of the inflaton 
is only bound below so that inflation continues at least 50 e-foldings.
We have calculated $\Delta$ under the various parameter sets which satisfy the CMB normalization 
and the results are shown in Fig.~\ref{Fig8}.
We found that there is no local minimum for $m_{3/2} \lesssim 100~\mathrm{TeV}$ $(n=2)$ or $m_{3/2} \lesssim 1000~\mathrm{TeV}$ $(n=4)$ for $f_a \sim 10^{15}~\mathrm{GeV}$.

From the above discussion, the local minimum of the inflaton potential may appear for large gravitino mass
and relatively small $n$, which leads to $\Delta < 1$.
In such a case, the initial values of $\varphi$ and $\theta_S$ are constrained in order to avoid trapping at the local minimum.
For successful inflation, one of the following conditions must be satisfied.
First, the initial value of $\theta_S$ must be small and the subsequent evolution must not change $\theta_S$ significantly in order to keep $\cos\theta_S > 0$ where local minimum of $\varphi$ does not appear.
Second, the initial value of $\varphi$ must be smaller than the local maximum given by
\be
	\varphi_\mathrm{max} = \frac{2^{(5n-1)/(6n-2)}f_a}{[2n(2n-1)]^{n/(3n-1)}} \bigg[ \frac{n \mu^2}{(2n-1)m_{3/2}f_a(-\cos \theta_S)} \bigg]^{(n-1)/(3n-1)},
\ee
provided $\cos\theta_S < 0$.
To investigate the allowed region of the initial value of the inflaton, we integrate the following equations of motion of the inflaton : 
\be
	3H \dot\varphi \simeq - \mu^4 \bigg[ \frac{4^{(2n-1)/(n-1)}}{[2n(2n-1)]^{n/(n-1)}} \frac{n}{2n-1} \bigg( \frac{f_a}{\varphi} \bigg)^{2n/(n-1)} \frac{1}{\varphi} + \frac{1}{4} \varphi^3 \bigg]
	-2 \sqrt{2} \mu^2 m_{3/2} \cos\theta_S,
\ee
\be
	3H \dot\theta_S \simeq \frac{2\sqrt{2}}{\varphi} \mu^2 m_{3/2} \sin \theta_S.
\ee

\begin{figure}[tp]
\centering
\subfigure[$n=2$ and $m_{3/2} = 100~\mathrm{TeV}$]{
	\includegraphics [width = 7.5cm, clip]{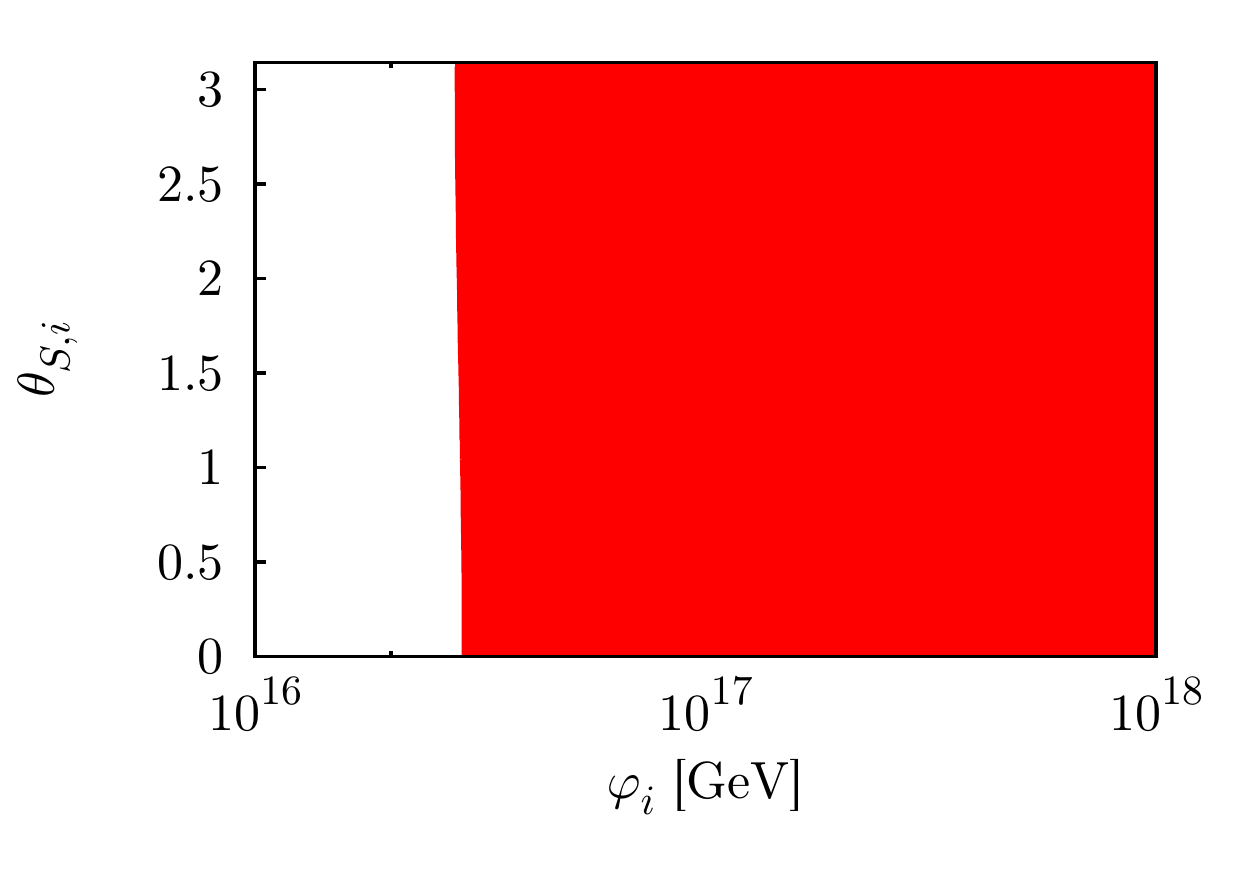}
	\label{Fig5a}
}
\subfigure[$n=2$ and $m_{3/2} = 1000~\mathrm{TeV}$]{
	\includegraphics [width = 7.5cm, clip]{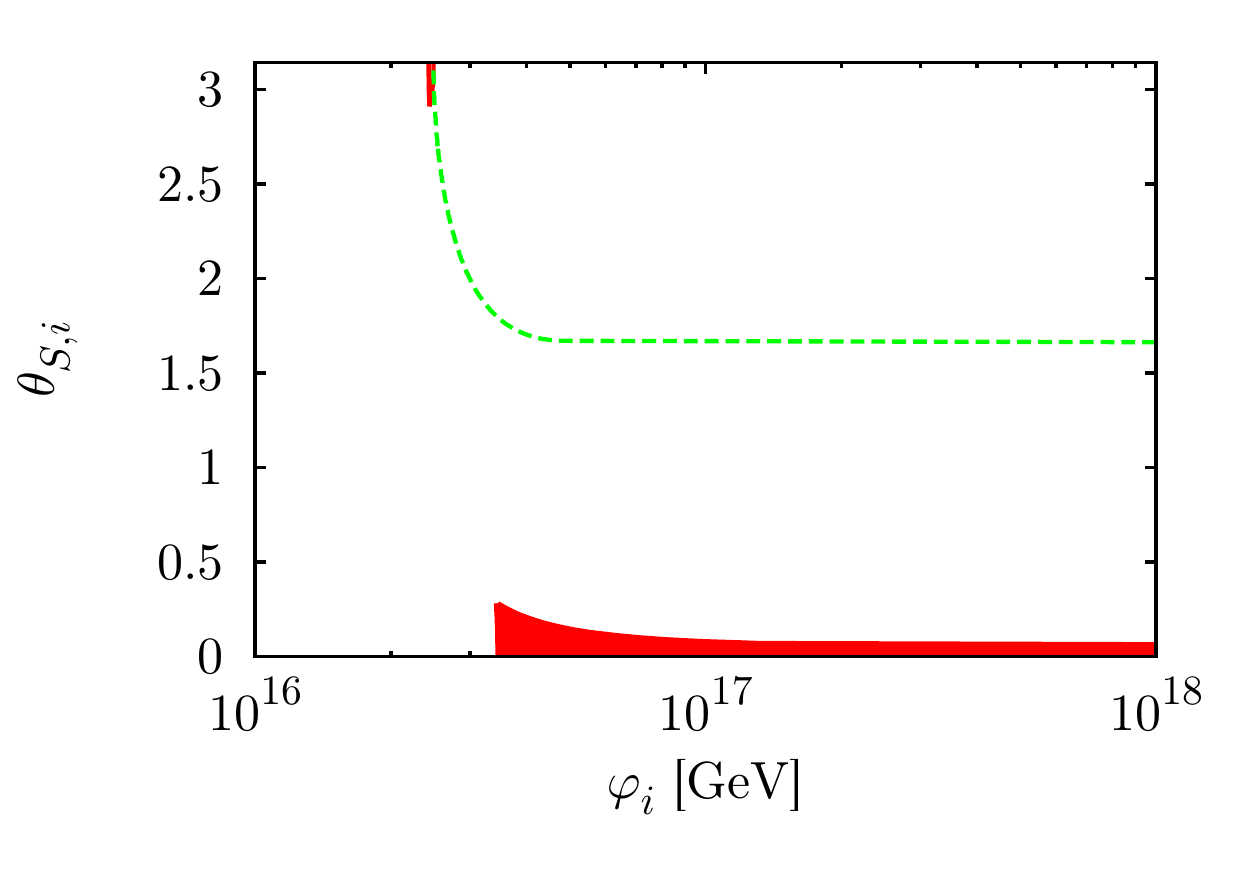}
	\label{Fig5b}
}
\caption{%
	The allowed regions of the initial value of the inflaton are shown as the red 
	regions.
	We have taken $f_a = 10^{15}~\mathrm{GeV}$, $\mu = 3 \times 10^{14}~\mathrm{GeV}$ and $n=2$. 
	The lower limit of $\varphi$ comes from the condition that the inflation continues 
	at least 50 e-foldings.
	In Fig.~\ref{Fig5b}, the dashed green line represennts $\varphi_\mathrm{max}$ 
	and the empty region corresponds to the breakdown of our assumption 
	$\theta_S \sim \mathrm{constant}$.
}
\label{Fig5}
\end{figure}

As a result, we show the allowed regions of the initial values of the inflaton field 
in Fig.~\ref{Fig5}.
We have taken $f_a = 10^{15}~\mathrm{GeV}$, $\mu = 3 \times 10^{14}~\mathrm{GeV}$ 
and $n=2$. 
From these results, there is no severe initial value problem in the smooth hybrid inflation model even if we adopt relatively large gravitino mass 
$m_{3/2} \sim 100~\mathrm{TeV}$, 
which is contrasted to the case of SUSY hybrid inflation~\cite{Nakayama:2010xf}.

Before ending this section, we comment on the advantages of the smooth hybrid inflation model compared with the hybrid inflation model~\cite{Kawasaki:2010gv,Kawasaki:2011ym}.
\begin{itemize}
	\item
	Because the PQ symmetry is already broken during inflation, the initial 
	misalignment angle at each spatial point takes an identical value at least 
	in our observable Universe, 
	and hence cosmic strings and domain walls are never formed after inflation.
	In the hybrid inflation model, on the other hand, the PQ symmetry is broken 
	at the end of inflation and the initial misalignment angles take random values 
	at different spacial points.
	In such a case, the cosmic strings are formed at the PQ symmetry breaking and 
	the domain walls are formed at the QCD phase transition~\cite{Sikivie:1982qv}, 
	which induces the additional contribution to the axion cold dark 
	matter~\cite{Hiramatsu:2010yu,Hiramatsu:2012gg}.
	\item
	In the smooth hybrid inflation model, an initial misalignment angle 
	at each spatial point can be chosen to be arbitrarily small value from the same 
	reason as above. 
	This implies the PQ symmetry breaking scale can be larger than the usual 
	upper limit $10^{12}\,\mathrm{GeV}$ without conflicting with the CDM density 
	parameter even without late-time entropy production.
	\item
	The scalar spectral index can be well consistent with the WMAP central value 
	without invoking tuning in the non-minimal K\"ahler potential.
	In the hybrid inflation model, on the other hand, the scalar spectral index is 
	relatively large $(n_s > 0.98)$, under the minimal K\"ahler potential.
	Although the WMAP central value can be reproduced under the non-minimal 
	K\"ahler potential~\cite{BasteroGil:2006cm,urRehman:2006hu}, a severe initial 
	value problem of the inflaton arises.
	These problems do not exist in the smooth-hybrid inflation model.
	\item
	The presence of the constant term in the superpotential is not so effective 
	compared with the SUSY hybrid inflation case.
	In the SUSY hybrid inflation model, the parameter space is strongly constrained 
	for the large gravitino mass under the minimal K\"ahler 
	potential~\cite{Nakayama:2010xf}.
	In the smooth hybrid inflation model, we found that, from Fig.~\ref{Fig2c} 
	and \ref{Fig2d}, the effects of the constant term is 
	small up to $m_{3/2} \sim 100-1000\,\mathrm{TeV}$ for most parameter regions.
	Note also that the angular motion of the inflaton field does not spoil 
	the inflaton dynamics for $m_{3/2} \lesssim 100-1000~\mathrm{TeV}$ 
	even if $\theta_S \simeq \pi/2$ 
	because the linear term in the inflaton potential can never dominate the dynamics, 
	hence there is no severe initial value problem of the inflaton in the smooth 
	hybrid inflation model for $m_{3/2} \lesssim 100~\mathrm{TeV}$.
	\item
	A drawback of the smooth hybrid inflation model is that it predicts the large axion 
	isocurvature perturbation, which poses severe constraint 
	on the case of $n=2$, as shown in Sec.~\ref{isocurvature}.	
\end{itemize}

\begin{figure}[t]
\centering
\subfigure[$n=2$]{
	\includegraphics [width = 7.5cm, clip]{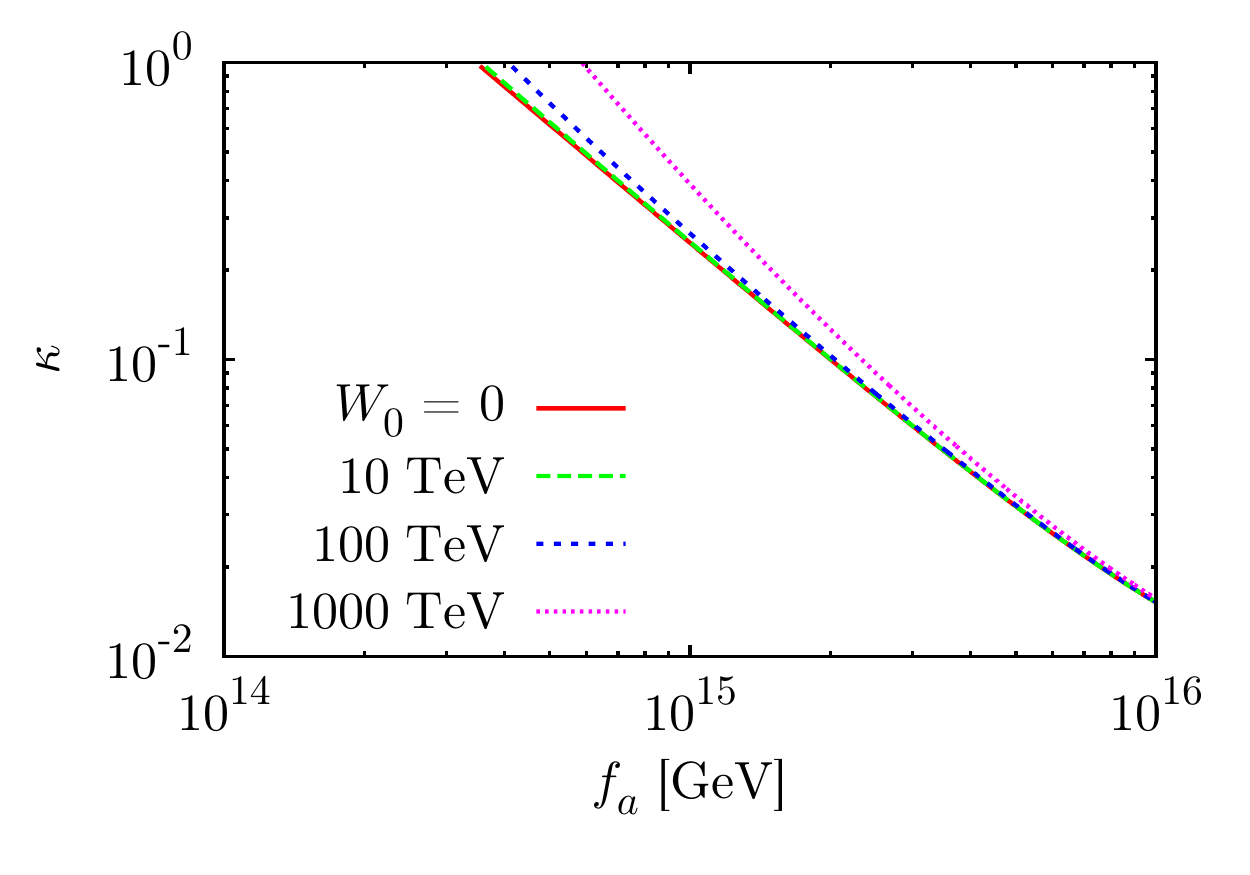}
	\label{Fig9a}
}
\subfigure[$n=4$]{
	\includegraphics [width = 7.5cm, clip]{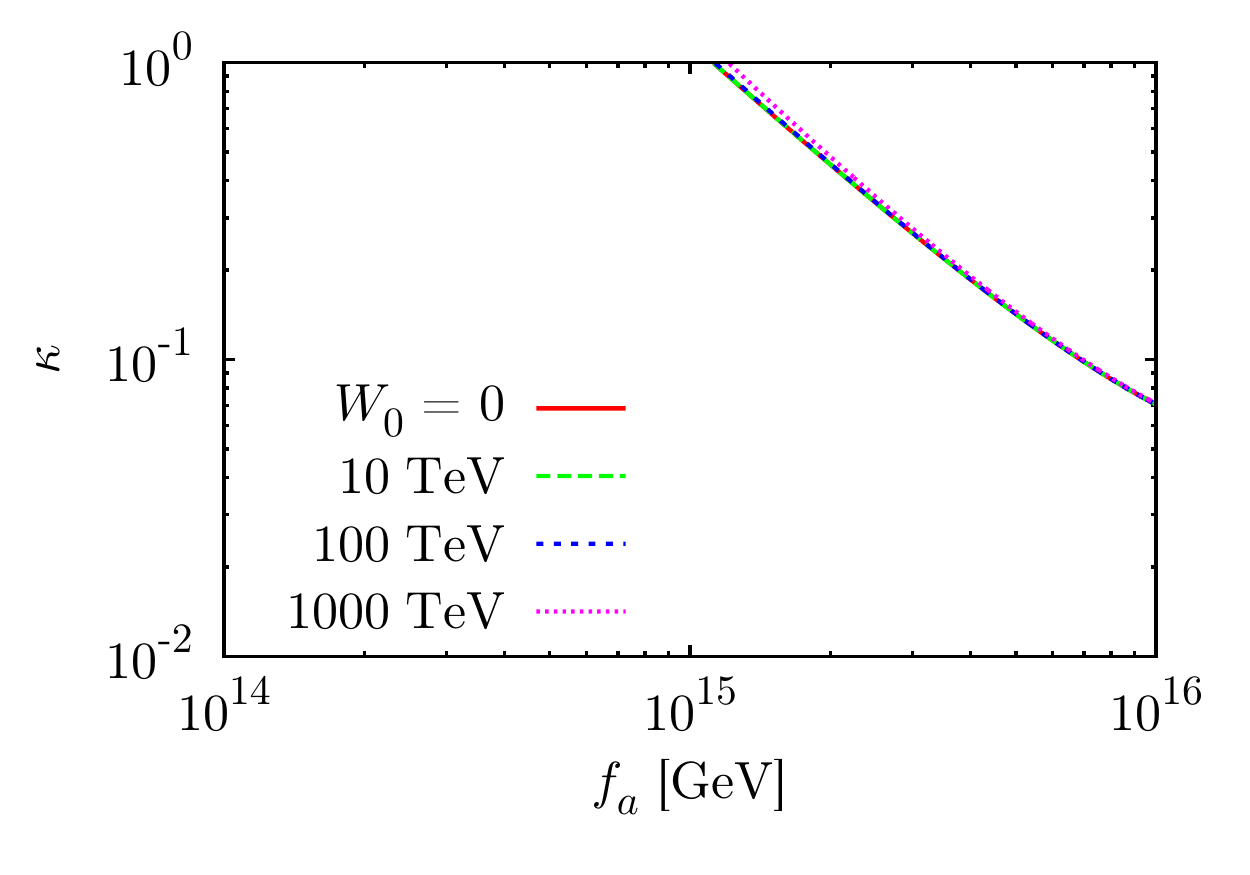}
	\label{Fig9b}
}
\caption{
	$\kappa$ as a function of $f_a$ after the WMAP 
	normalization of the density perturbation is imposed.
	We have taken $n=2$ in Fig,~\ref{Fig9a} and $n=4$ in Fig.~\ref{Fig9b} and $W_0=0$ (solid-red lines), $m_{3/2}=10~\mathrm{TeV}$ (dashed-green lines), $m_{3/2}=100~\mathrm{TeV}$ 
	(dotted-blue lines) and $m_{3/2}=1000~\mathrm{TeV}$ (small-dotted-magenta line) and
	the e-folding number is set to be 50 in both figures.
}
\label{Fig9}
\end{figure}

\section{The scalar dynamics after inflation and cosmology} 
\label{dynamics}

\subsection{Reheating after inflation}

In this section, we consider the PQ field dynamics after inflation.
The post-inflationary dynamics of the PQ fields is determined by 
the following low-energy potential:
\be
	\begin{split}
	V &= \bigg| - \mu^2 + \frac{ (\Psi \bar\Psi)^n }{ M^{2(n-1)} } \bigg|^2 
	+ \frac{ n^2 |S|^2 | \Psi \bar\Psi |^{2(n-1)} }{ M^{4(n-1)} } 
	( |\Psi|^2 + | \bar\Psi |^2 ) \\[2mm]
	&~~~ + 2 m_{3/2}\bigg[ S \bigg( \mu^2 +(n-1) \frac{(\Psi \bar\Psi)^n}{M^{2(n-1)}}\bigg) + \mathrm{c.c.} \bigg]
	+ m_{3/2}^2 (|S|^2 + |\Psi|^2 + |\bar\Psi|^2-3)
	\end{split}
	\label{pot_after_inf}
\ee
Right after the inflation, the low-energy SUSY breaking terms written in the second row in (\ref{pot_after_inf}) are negligible 
and the minima of the PQ fields are given by $v \approx \bar v \approx f_a$.
The flat direction which we identify as the saxion ($\sigma$) 
and the direction perpendicular to the flat direction which is denoted by $\Phi$ 
are related to the PQ scalar fields as
\be
	|\Psi| = v \bigg( 1 + \frac{\Phi + \sigma}{\sqrt{2} F_a} \bigg),
	~~~|\bar\Psi| = \bar v \bigg( 1 + \frac{\Phi - \sigma}{\sqrt{2} F_a} \bigg)
\ee
From the above relation, the first term in the right-hand-side 
of (\ref{pot_after_inf}) gives the mass of $\Phi$ and the second term 
gives the mass of $S$.
$\Phi$ and $S$ have the same mass : $m_\Phi = m_S = \sqrt{2}\kappa f_a$ 
for $v=\bar v = f_a$ where $\kappa = n (f_a/M)^{2(n-1)}$ which takes a value 
shown in Fig.~\ref{Fig9}.
Furthermore they have the same initial amplitude of order $f_a$, so they equally 
contribute to the total energy density of the universe.
Because of the large mass and large coupling constant of $\Phi$ and $S$, 
their lifetimes are quite short.

We must note the mixing of $S$ and $\Phi$.
Redefining $\varphi$ as $\mathrm{Re}(S)/\sqrt{2}$, the mixing term is written as $4n(n-1)m_{3/2} \mu^2 \varphi \Phi/F_a \sim m_{3/2} m_S \varphi \Phi$ in the potential (\ref{pot_after_inf}).
Because of $m_S = m_\Phi$, the mass eigenstates are $(\varphi \pm \Phi)/\sqrt{2}$ whose squred masses are given by $m_S^2 \pm \delta m^2$ respectively, where $\delta m^2 \sim m_{3/2}m_S$.
Hence the mixing time scale is estimated as $2m_S / \delta m^2 \sim 1/ m_{3/2}$.
This is much longer than the lifetime of both $S$ and $\Phi$ which is calculated below, so we can neglect the mixing effect here.

Let us first consider the decay of $S$.
From the superpotential (\ref{superpot_axion}), $S$ has a Yukawa interaction with the axino.
Representing the fermionic components of the PQ superfields $\Psi$ ($\bar\Psi$) as $\psi$ ($\bar\psi$) respectively,
the axino field is defined as $\tilde a = (\psi-\bar\psi)/\sqrt 2$, whose mass is given by $\kappa |v_S| (=m_{3/2}/n)$,
where $v_S=\langle S\rangle = -2m_{3/2}\mu^2/m_S^2=-m_{3/2}/(\kappa n)$ is the VEV of $S$.
The other combinations of $\tilde S$ (fermionic components of $S$) and $ (\psi+\bar\psi)/\sqrt 2$ get masses of $\simeq \sqrt{2}\kappa f_a$.
The interaction of the axino with $S$ is derived from the superpotential (\ref{superpot_axion}) as
\be
	-\mathcal L_{S\text{-}\tilde a} = \frac{1}{2} \kappa S \tilde a \tilde a + \mathrm{h.c.}.
\ee
From this term, the decay rate of $S$ into axino pair is derived as
\be
	\Gamma_{S \to \tilde a \tilde a} = \frac{\kappa^2 m_S}{32\pi} = \frac{\sqrt 2}{32 \pi} \kappa^3 f_a.
\ee
In the present model, the coupling constant is relatively large such as 
$\kappa \sim 0.1$~--~1, 
which implies the axino production occurs soon after inflation.
The decay temperature of $S$ defined through the axino energy density is calculated as
\be
	T_{\tilde a}(t_S) = \bigg( \frac{\pi^2}{90} g_{\tilde a} \bigg)^{-1/4} \sqrt{\Gamma_{S \to \tilde a \tilde a} M_P}
	\simeq 3 \times 10^{14}\,\mathrm{GeV} \bigg( \frac{\kappa}{0.1} \bigg)^{3/2} \bigg( \frac{f_a}{10^{15}\,\mathrm{GeV}} \bigg)^{1/2}
\ee
where $g_{\tilde a} = 2 \times 7/8 = 1.75$ is the relativistic degrees of freedom of axinos and $t_S$ represents the time of the $S$ decay.
Since the axino decoupling temperature is given by $T_D \sim 10^{17}\,\mathrm{GeV}(f_a / 10^{15}\,\mathrm{GeV})^2$~\cite{Rajagopal:1990yx}, the produced axinos cannot have any thermal contacts with the MSSM particles.
For convenience, however, we call $T_{\tilde a}$ as the decay temperature of $S$.

Now we consider the decay of $\Phi$.
Similar to the saxion, $\Phi$ decays mainly into gluons and gluinos through $X$ and $\bar X$ loops, 
so the Universe is reheated by the $\Phi$ decay.
The decay rate of $\Phi$ into gluons is calculated as
\be
	\Gamma_{\Phi \to gg} \simeq \frac{\alpha_s^2}{64\pi^3} \frac{m_{\Phi}^3}{F_a^2} 
	\simeq\frac{\sqrt{2}\alpha_s^2}{64 \pi^3} \kappa^3 f_a.
\ee
Thus the reheating temperature is given by
\be
	T_R \simeq \bigg( \frac{\pi^2}{90} g_* \bigg)^{-1/4} \sqrt{\Gamma_{\Phi \to gg} M_P} 
	\simeq 1 \times 10^{12}\,\mathrm{GeV} \bigg( \frac{\alpha_s}{0.04} \bigg) 
	\bigg( \frac{\kappa}{0.1} \bigg)^{3/2} \bigg( \frac{f_a}{10^{15}\,\mathrm{GeV}} \bigg)^{1/2},
\ee
where $g_*$ is the relativistic degree of freedom which we take $g_* \approx 100$ in the last equality.
After the decay of $S$ into axinos until the decay of $\Phi$, the universe is dominated by $\Phi$ because the produced axino from $S$ is highly relativistic 
and its energy density is redshifted relative to that of $\Phi$ as $\rho_{\tilde a} \propto a^{-1} \rho_{\Phi}$.
Thus we can calculate the axino abundance as 
\be
	\frac{n_{\tilde a}}{s} \bigg|_{\Phi} = \frac{3 T_R}{2 m_S} 
	\simeq 0.01 \bigg( \frac{0.1}{\kappa} \bigg) \bigg( \frac{10^{15}\,\mathrm{GeV}}{f_a} \bigg) \bigg( \frac{T_R}{10^{12}\,\mathrm{GeV}} \bigg), 
	\label{axino_abun}
\ee
which implies the overproduction of axinos, but the late-time entropy production due to the saxion decay dilutes it.
Cosmological implications will be discussed below.

\subsection{Saxion dynamics and its implications}

After the decay of $\Phi$, the Universe is radiation dominated.
The existence of thermal background induces thermal effects on the scalar potential.
In particular, in the KSVZ model, $\Psi$ interacts with heavy quarks $Q$ and $\bar Q$ and heavy quarks interact with with MSSM particles through the QCD couplings.
Then the two-loop correction to the strong coupling constant induces the thermal effective potential~\cite{Anisimov:2000wx} such as
\be
	V_\mathrm{th} \simeq \alpha_s^2 T^4 \log \frac{|\Psi|^2}{T^2}.
	\label{thermal-log}
\ee
The saxion dynamics is similar to the previous studies~\cite{Kawasaki:2010gv,Kawasaki:2011ym}, 
and here we briefly repeat the discussion.
This thermal-log potential lifts the flat direction, and $|\Psi|$ ($|\bar\Psi|$) tends to roll down to the smaller (larger) value.
Just after the reheating, both $\Psi$ and $\bar\Psi$ are placed at $f_a$ and the thermal mass is larger than the Hubble parameter.
Thus $|\bar\Psi|$ rolls down towards a larger value without feeling the Hubble friction.
The thermal mass decreases as $\bar\Psi$ rolls down and becomes comparable to the Hubble parameter and then $\bar\Psi$ gets frozen due to the Hubble overdamping at the field value of $|\bar\Psi| \sim \alpha_s M_P$.
As the Universe expands and temperature decreases, the thermal effect becomes insignificant and the zero-temperature minimum (\ref{minimum}) appears. At this epoch, the PQ field is displaced from the zero-temperature minimum.
As a result, when the Hubble parameter becomes comparable to the gravitino mass, the PQ field starts to oscillate around its true minimum along the flat direction, with an amplitude of $\sigma_i \sim \alpha_s M_P$. 

\begin{figure}[t]
\centering
\includegraphics [width = 10cm, clip]{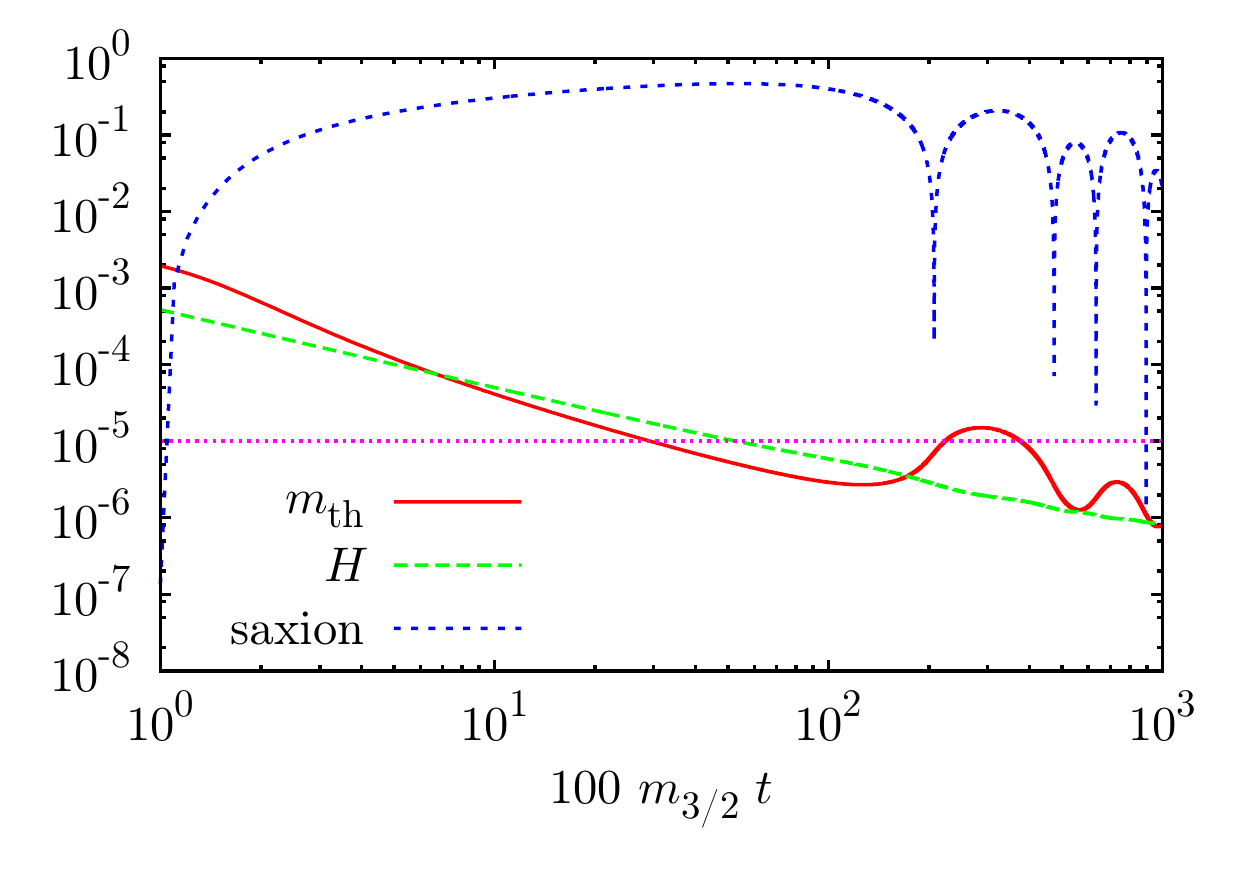}
\caption{
	The dynamics of the saxion after reheating are shown.
	We show the time evolution of $m_\mathrm{th}$ (red solid line), 
	Hubble parameter (green dashed line) and saxion (blue dotted line).
	We have taken $\mu = 0.05$,  $M = 0.1$ ($f_a = 0.07$), $\alpha_s=1$, 
	$m_{3/2} = 10^{-5}$, $c_1 = 1$ and $c_2 = 2$ in units of $M_P = 1$ 
	and $S = 0.1f_a$ and $\Psi = \bar\Psi = f_a$ as initial values.
	The small-dotted magenta line represents the gravitino mass.
}
\label{Fig4}
\end{figure}

In Fig.~\ref{Fig4} we show the above mentioned dynamics by numerically solving the equation of motions of fields.
In this figure, the time evolution of effective mass 
$m_\mathrm{th} (\simeq \alpha_s T^2/|\bar\Psi|)$ (red solid line), 
Hubble parameter (green dashed line) and saxion field value (blue dotted line) are shown.
The saxion energy density divided by the entropy density after the oscillation is thus given by
\begin{equation}
	\left.\frac{\rho_\sigma}{s}\right|_{\rm osc}=\frac{1}{8}T_{\rm osc}\left( \frac{\sigma_i}{M_P} \right)^2
	\sim 4\times 10^7\,{\rm GeV}\left( \frac{m_\sigma}{100\,{\rm TeV}} \right)^{1/2}
	\left( \frac{\sigma_i}{\alpha_s M_P} \right)^2.
\end{equation}
Note that the entropy density in this expression includes only that produced by the inflaton decay.
Since the saxion dominates the Universe before it decays, its decay dilutes the abundance of preexisting matter.
The dilution factor is given by
\begin{equation}
	\gamma \equiv \frac{3T_\sigma}{4}\left( \frac{\rho_\sigma}{s} \right)^{-1}
	\sim 2\times 10^{-9}	\left( \frac{T_\sigma}{100\,{\rm MeV}} \right)
	\left( \frac{100\,{\rm TeV}}{m_\sigma} \right)^{1/2}
	\left( \frac{\alpha_s M_P}{\sigma_i} \right)^2.
\end{equation}
Number densities per entropy density of all matters, which are produced at the reheating 
and conserved thereafter such as the gravitino and axino, are diluted by this factor.
As a result, thermally produced gravitinos~\cite{Bolz:2000fu} as well as the non-thermally produced ones by the direct inflaton decay
\cite{Endo:2006zj,Kawasaki:2006gs} are also diluted away to a cosmologically negligible level.
Thermally produced axino abundance during the reheating~\cite{Covi:2001nw,Chun:2011zd} is also negligibly small due to the large PQ scale and huge dilution.

On the other hand, the axino abundance produced by the inflaton decay, given by Eq.~(\ref{axino_abun}), may not be neglected
even after the dilution by the saxion decay.
The axino obtains a mass of $\simeq m_{3/2}/n$ and its decay width into the gluino and gluon is given by
\begin{equation}
	\Gamma_{\tilde a  \to \tilde g g} \simeq \frac{\alpha_s^2}{32\pi^3}\frac{m_{\tilde a}^3}{F_a^2}.
\end{equation}
Since the decay width is comparable to the saxion in the KSVZ model, the LSPs produced by the axino decay
is always smaller than or comparable to those produced from the saxion decay.\footnote{
	In the DFSZ model, the axino decay rate into Higgs boson and higgsino is comparable to the
	saxion decay rate into the higgsino pair (\ref{sigma-higgsino}).
}
Below we examine the LSP abundance from the saxion decay, taking account of self-annihilation of the LSP.

\subsection{Wino dark matter from saxion decay} \label{Wino}


In this section we see that Winos produced by the saxion decay can account 
for the present dark matter abundance.\footnote{
	In the next section, we will see that the axion cannot be the dominant component of DM once 
	the isocurvature constraint is taken into account.
}
Assuming the anomaly mediated SUSY breaking (AMSB) model~\cite{Randall:1998uk,Giudice:1998xp},
or the pure-gravity mediation model~\cite{Ibe:2011aa} in which the 125\,GeV Higgs boson can easily be explained,
the gravitino mass is of the order of $\sim 10^2$~--~$10^3\,\mathrm{TeV}$.
The AMSB contributions to the the MSSM gaugino masses, $M_a = (M_1, M_2, M_3)$, are given by 
\be
	M_a = \frac{b_a}{16 \pi^2} g_a^2 m_{3/2},
\ee
where $g_a$ are the SM gauge coupling constants and $b_a = (11, 1, -3)$.
Then, the ratio of the MSSM gaugino masses are given by 
$m_{\tilde B} : m_{\tilde W} : m_{\tilde g} \simeq 3: 1: 8$ implying the Wino-LSP 
with mass of $O(100)\,\mathrm{GeV}$~--~$O(1)\,\mathrm{TeV}$,
although the relation may be modified by the Higgs-higgsino loop contribution.
Because of the large gravitino mass, the saxion decay temperature given 
by (\ref{T_sigma_K}) or (\ref{T_sigma_D}) becomes $O(1)\,\mathrm{GeV}$.
In such a case, the present dark matter abundance can be explained 
by the Wino-LSP from the saxion decay as shown below.

Because we consider the large gravitino mass of order $O(100)\,\mathrm{TeV}$ 
and the LSP mass is given by $m_{\chi} \sim m_{3/2} / 400$, 
the freeze-out temperature of the LSP, given by $T_\mathrm{fr} \approx m_\chi / 25$, 
may be higher than the saxion decay temperature, which implies the produced LSPs 
are never in thermal equilibrium.
Hence, in the following arguments, we focus on the non-thermally produced 
dark matter~\cite{Moroi:1999zb,Fujii:2001xp,Endo:2006ix}.

Since the decay rate of the saxion into SUSY particles is comparable to that into ordinary particles, 
(see (\ref{sigma-gg}) and (\ref{sigma-gluino}) in the KSVZ model or (\ref{sigma-hh}) and (\ref{sigma-higgsino}) in the DFSZ model) 
a large number of LSPs are produced. 
The annihilation cross section of Wino-LSP is given by~\cite{Moroi:1999zb}
\be
	\langle \sigma_\mathrm{ann} v \rangle = \frac{g_2^4}{2 \pi} \frac{1}{m_\chi^2} \frac{(1-x_W)^{3/2}}{(2-x_W)^2},
\ee
where $x_W = m_W^2 / m_\chi^2$ with $m_W$ being the $W$ boson mass.
This is roughly estimated as $\langle \sigma_\mathrm{ann} v \rangle \sim 10^{-7}\,\mathrm{GeV}^{-2}$ for $m_{\chi} \sim 100\,\mathrm{GeV}$. 
Note that the Wino cross section is constrained from WMAP 
observations~\cite{Slatyer:2009yq,Kanzaki:2009hf}
and recent gamma-ray measurements by the Fermi satellite~\cite{Hooper:2012sr},
which excludes the Wino mass below $\sim 300$\,GeV as a dominant component dark matter.
Note that we have ignored the Sommerfeld enhancement effect~\cite{Hisano:2003ec}. 
This is valid as long as we consider the Wino lighter than 
$\sim 1$\,TeV~\cite{Hisano:2006nn}.

\begin{figure}[t]
\centering
\includegraphics [width = 10cm, clip]{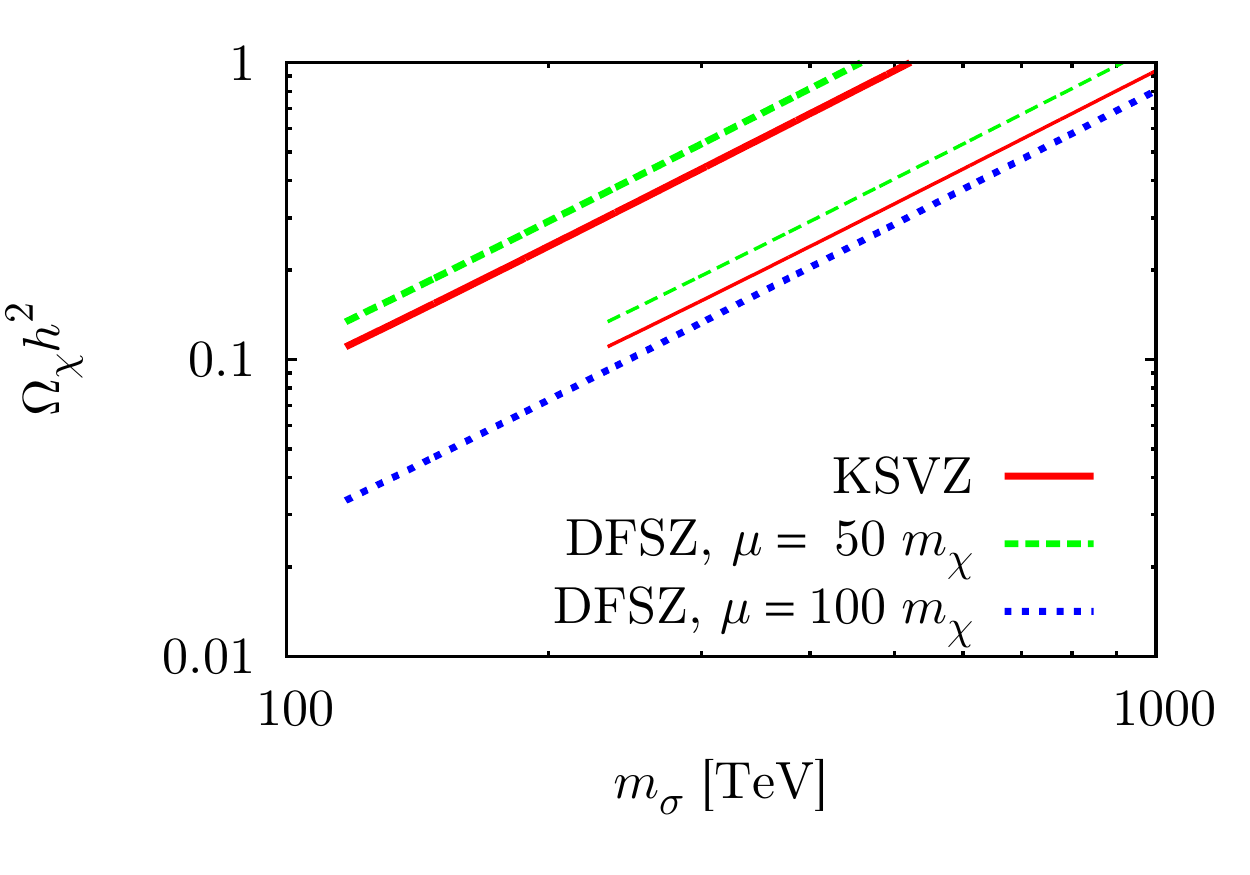}
\caption{
	The relations between the wino abundance and the saxion mass are shown.
	The solid red lines, the dashed green lines and dotted blue line correspond to the KSVZ model, the DFSZ model with $\mu = 50m_\chi$ (higgsino mass) and the DFSZ model with $\mu = 100m_\chi$ respectively.
	We have taken $m_\sigma = m_{3/2}$ (thick lines) and $m_\sigma = 2m_{3/2}$ (thin lines).
	The breakiing points reflect the constraint on the Wino mass, $m_\chi > 300~\mathrm{GeV}$.
}
\label{Fig6}
\end{figure}

The abundance of the LSP is determined by solving the following Boltzmann equation: 
\be
	\frac{dn_{\chi}}{dt} + 3 H n_{\chi} = - \langle \sigma_\mathrm{ann} v \rangle n_{\chi}^2,
\ee
where $n_\chi$ is the LSP number density.
Assuming $\langle \sigma_\mathrm{ann} v\rangle$ is temperature-independent, this is analytically solved and we get 
the LSP abundance in terms of $Y_{\chi}(T) = n_{\chi}(T) / s$ as
\be
	Y_{\chi}^{-1}(T) = Y_{\chi}^{-1}(T_\sigma) + \bigg( \frac{8\pi^2}{45}g_*(T_\sigma) \bigg)^{1/2} \langle \sigma_\mathrm{ann} v \rangle M_P (T_\sigma - T),
	\label{chi_yield}
\ee
where $Y_{\chi}(T_\sigma)$ contains the contributions from both non-thermally produced LSPs by the saxion and axino decay
and also thermal freeze-out LSPs.
From this expression, we soon realize that the result is independent of the initial abundance $Y_{\chi}(T_\sigma)$ 
if the initial abundance or the annihilation cross section is large enough, as is expected.
Actually, in the present setup, the abundance of LSPs from the saxion decay is large enough to annihilate efficiently.
Thus the resultant density parameter is estimated as 
\be
	\Omega_{\chi} h^2 \simeq 0.08 \bigg( \frac{60}{g_*(T_\sigma)} \bigg)^{1/2} \bigg( \frac{m_{\chi}}{100\,\mathrm{GeV}} \bigg) 
	\bigg( \frac{10^{-8}\,\mathrm{GeV}^{-2}}{\langle \sigma_\mathrm{ann} v \rangle} \bigg) \bigg( \frac{1\,\mathrm{GeV}}{T_\sigma} \bigg).
\ee
This implies that the non-thermally produced Winos can eventually becomes the dominant component of the dark matter.
Taking into account the cosmological and astrophysical constraint on the Wino mass, $m_\chi > 300~\mathrm{GeV}$~\cite{Ibe:2011aa}, the resultant abundance of the Wino dark matter is shown in Fig.~\ref{Fig6}.
From this, the mass of the saxion is predicted as $m_\sigma \sim 100~{\rm TeV}$ in both KSVZ and DFSZ axion models with $\mu > 50m_\chi$.
Such a Wino LSP may be found by direct or indirect detection experiments or LHC~\cite{Moroi:2011ab,Bhattacherjee:2012ed}.

\section{Isocurvature perturbation and non-Gaussianity} \label{isocurvature}

In this section, we calculate the axion abundance and the constraints from the isocurvature perturbation.
We also calculate the non-Gaussianity from the isocurvature perturbation.

We first note that the axion cannot be diluted by the saxion decay once we impose $T_\sigma \gtrsim 1$\,GeV
in order to realize the Wino dark matter as shown above. 
Hence we have to tune the initial misalignment angle to suppress the axion abundance.
The present density parameter of the axion is given by~\cite{Gross:1980br,Turner:1985si,Lyth:1991ub}
\be
	\Omega_a h^2 \simeq 
	\begin{cases}
		0.2 \bigg(\cfrac{f_a}{10^{12}\,\mathrm{GeV}} \bigg)^{1.18} \theta_i^2 ~~~&\text{for}~~~\theta_i > \cfrac{H_I}{2\pi|\Psi(k_0)|} \\[4mm]
		0.2 \bigg(\cfrac{f_a}{10^{12}\,\mathrm{GeV}} \bigg)^{-0.82} \bigg(\cfrac{H_I f_a / 2 \pi|\Psi(k_0)|}{10^{12}\,\mathrm{GeV}} \bigg)^2~~~&\text{for}~~~\theta_i < \cfrac{H_I}{2\pi|\Psi(k_0)|},
	\end{cases}
	\label{axion_abun}
\ee
which means that the axion abundance can no longer be reduced by tuning $\theta_i$ if $\theta_i$ is smaller than the critical value given by 
\be
	\theta_\mathrm{cr} = \frac{H_I / 2\pi}{|\Psi(k_0)|} \simeq 2 \times 10^{-4} \bigg( \frac{H_I / |\Psi(k_0)|}{10^{-3}} \bigg),
\ee
where $\Psi(k_0)$ represents the PQ field value when the pivot scale leaves the horizon during inflation.\footnote{
Since $H/\Psi$ decreases during inflation, the critical value $\theta_{\rm cr}$ 
should be evaluated when the largest observable scale ($\sim k_0$) leaves the horizon.}

\subsection{Constraint from isocurvature perturbation} \label{isocurv}

Since the PQ symmetry is broken during inflation, there exists the quantum fluctuation of the axion, so the large CDM isocurvature perturbation can be induced from the axion 
and the model parameters are strongly constrained~\cite{Bean:2006qz,Kawasaki:2007mb}.
The ratio of the adiabatic curvature perturbation to the isocurvature one
is parameterized as
\be
	\frac{\mathcal{P}_\mathcal{S}}{\mathcal{P}_\mathcal{\zeta}} = \frac{\alpha}{1-\alpha},
\ee
where $\mathcal{P}_\zeta$ and $\mathcal{P}_\mathcal{S}$ are dimensionless power spectrum of the curvature and CDM 
isocurvature perturbations,
and $\alpha$ is constrained from the observation as $\alpha < 0.077$~\cite{Komatsu:2010fb}.
The power spectrum of the CDM isocurvature perturbation is related to that of the axion isocurvature perturbation in terms of the density parameters as
\be
	\mathcal{P}_\mathcal{S} = \bigg( \frac{\Omega_a}{\Omega_\mathrm{CDM}} \bigg)^2 \mathcal{P}_{\mathcal{S},a} 
	\label{cdm_iso}
\ee
and the power spectrum of the axion isocurvature perturbation is calculated as 
\be
	\mathcal P_{\mathcal{S},a}^{1/2} \simeq
	\begin{cases}
		\cfrac{H_I}{\pi |\Psi(k_0)| \theta_i} ~~~&\text{for}~~~\theta_i > \cfrac{H_I}{2\pi|\Psi(k_0)|}, \\[4mm]
		\cfrac{1}{4} ~~~&\text{for}~~~\theta_i < \cfrac{H_I}{2\pi|\Psi(k_0)|}
	\end{cases}
	\label{axion_iso}
\ee
 and 

\begin{figure}[t]
\centering
\subfigure[]{
	\includegraphics [width = 7.5cm, clip]{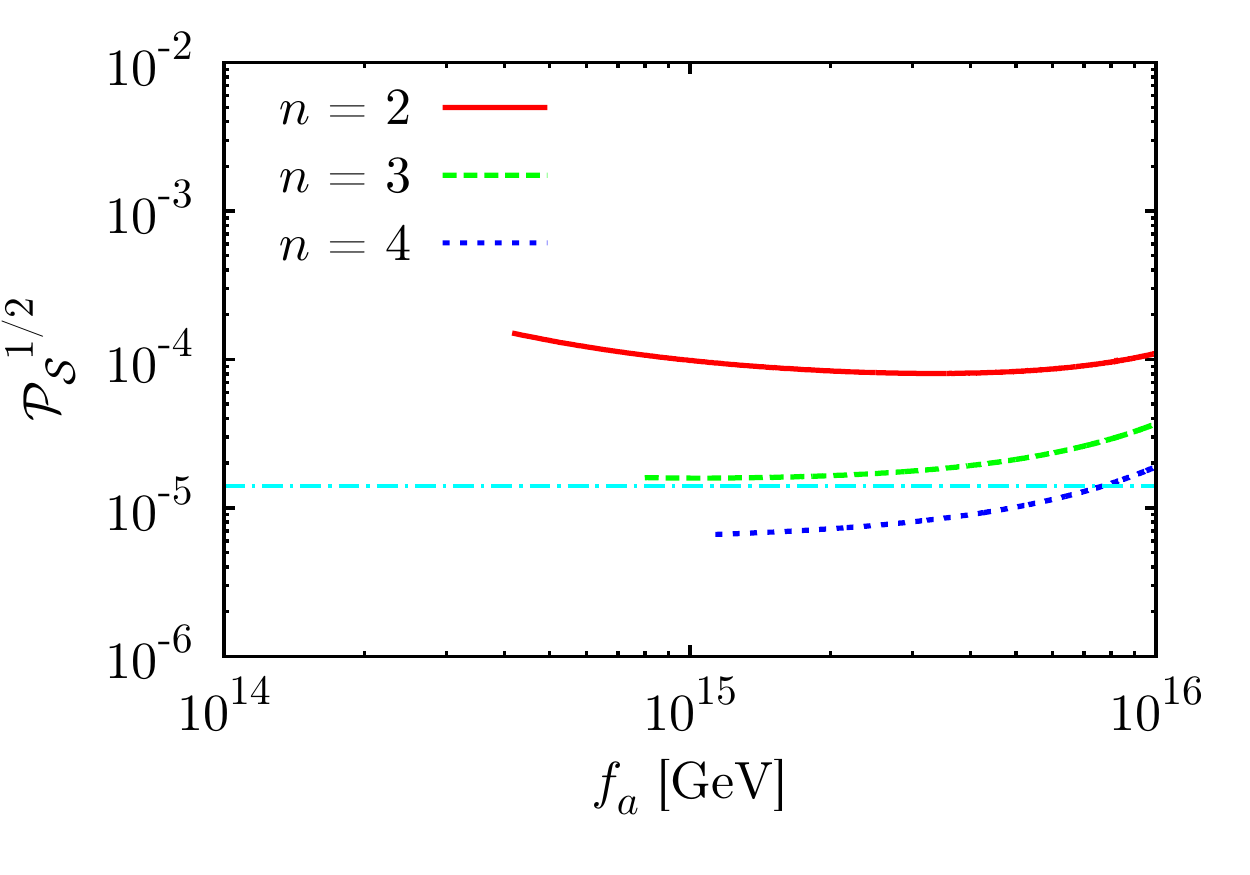}
	\label{Fig7a}
}
\subfigure[]{
	\includegraphics [width = 7.5cm, clip]{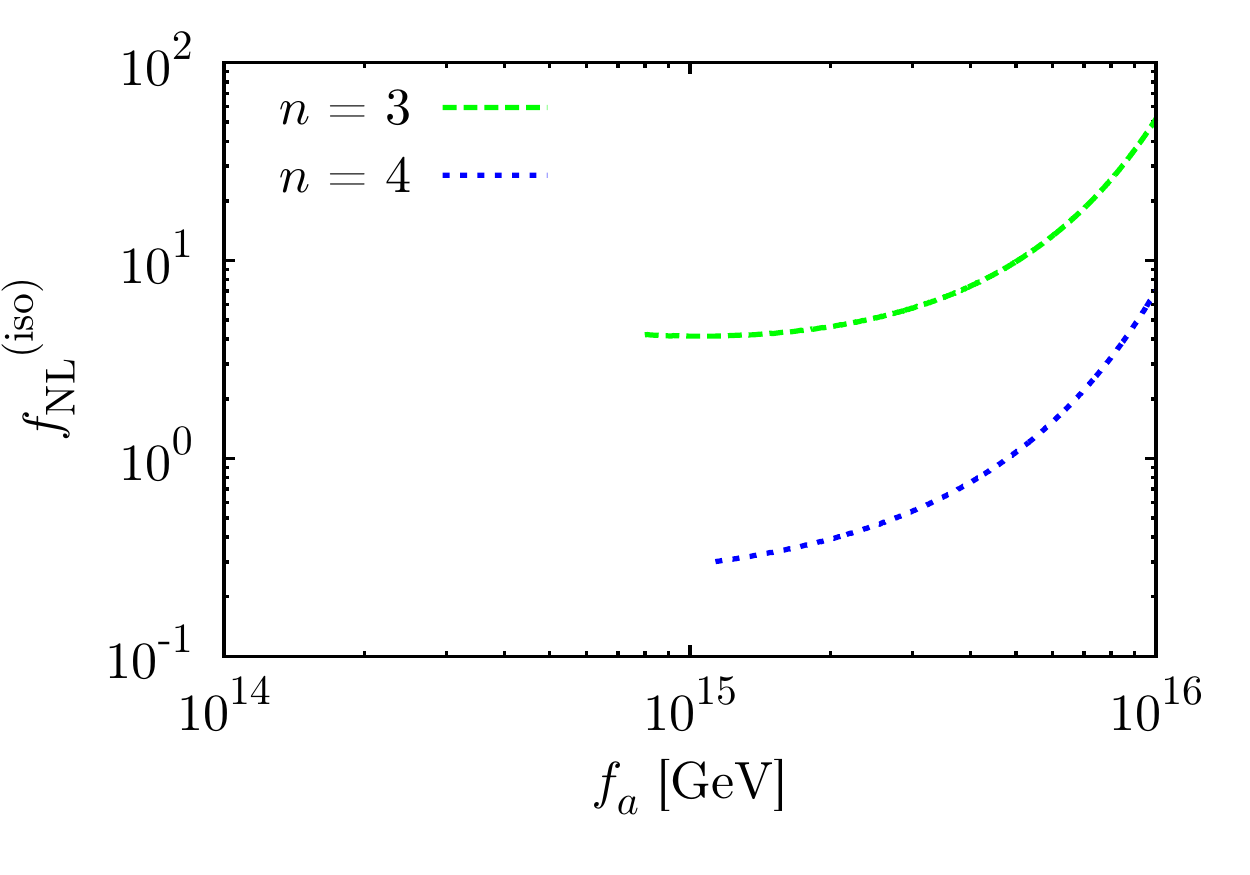}
	\label{Fig7b}
}
\caption{
	The power spectrum of the isocurvature perturbation (Fig.~\ref{Fig7a}) 
	and the non-linearity parameter (Fig.~\ref{Fig7b}) are shown.
	In both figures, we have taken $m_{3/2} = 100~\mathrm{TeV}$ 
	and $n=2$ (solid red line), $n=3$ (dashed green line) and $n=4$ (dotted blue line).
	In the left figure, the dashed-and-dotted cyan line represents 
	the upper limit from the observation.
}
\label{Fig7}
\end{figure}

Substituting (\ref{axion_abun}) and (\ref{axion_iso}) into (\ref{cdm_iso}) 
and using the WMAP best fit values $\mathcal P_{\zeta} = 2.43 \times 10^{-9}$ 
and $\Omega_\mathrm{CDM}h^2 \simeq 0.112$~\cite{Komatsu:2010fb}, 
the power spectrum the CDM isocurvature perturbation is rewritten as 
\be
	\mathcal{P}_\mathcal{S}^{1/2} = 
	\begin{cases}
	2 \theta_i \bigg( \cfrac{f_a}{10^{15}~\mathrm{GeV}} \bigg)^{1.18} \bigg( \cfrac{H_I / |\Psi(k_0)|}{10^{-3}} \bigg) ~~~&\text{for}~~~\theta_i > \cfrac{H_I}{2\pi|\Psi(k_0)|} \\[4mm]
	4 \times 10^{-5} \bigg( \cfrac{f_a}{10^{15}~\mathrm{GeV}} \bigg)^{1.18} \bigg( \cfrac{H_I / |\Psi(k_0)|}{10^{-3}} \bigg)^2~~~&\text{for}~~~\theta_i < \cfrac{H_I}{2\pi|\Psi(k_0)|}
	\end{cases}
	\label{P_iso}
\ee
This implies obviously that $\theta_i > H_I/ 2\pi |\Psi(k_0)|$ is not allowed by the constraint from the observation and the small initial misalignment angle satisfying $|\Psi(k_0)| \theta_i < H_I/ 2\pi$ is necessary for the case of $f_a \sim 10^{15}\,\mathrm{GeV}$. 
As seen from Eq.~(\ref{axion_abun}), the axion cannot be the dominant component of the CDM.
The constraint on the parameters are shown in Fig.~\ref{Fig7a}, in which $\theta_i < H_I / 2\pi |\Psi(k_0)|$ is assumed.
It is found that $n \leq 3$ is forbidden by the observation but $n=4$ is allowed for some range of $f_a$ of our interest.

\subsection{Non-Gaussianity from isocurvature perturbation} \label{NG}

In the inflationary paradigm, the primordial density perturbation, 
which is originated from the quantum fluctuation of the scalar field, 
obeys the almost Gaussian statistics.
Thus if the deviation from the Gaussianity is detected, it has rich information 
on the origin of primordial density perturbation.
In particular, the axion model can generate sizable non-Gaussianity in the isocurvature perturbation~ \cite{Boubekeur:2005fj,Kawasaki:2008sn}.
The non-Gaussianity is characterized by the non-linearity parameter $f_\mathrm{NL}$ defined by
\be
	\langle \zeta(\vec{k}_1) \zeta(\vec{k}_2) \zeta(\vec{k}_3) \rangle = \frac{6}{5} f_\mathrm{NL} (2 \pi)^3 \delta^{(3)}(\vec{k}_1 + \vec{k}_2 + \vec{k}_3) 
	[ P_\zeta (k_1) P_\zeta (k_2) + \text{(2 perms)} ],
\ee
where `2 perms' means 2 permutations.
$\zeta$ denote the curvature perturbation evaluated on the uniform density slicing and $P_\zeta(k)$ is the power spectrum of $\zeta$ defined as
\be
	\langle \zeta(\vec{k}) \zeta(\vec{k}') \rangle = (2 \pi)^3 P_\zeta(k) \delta^{(3)} (\vec{k} + \vec{k}').
\ee

In the present model, the isocurvature perturbation from the axion becomes large and the axion density parameter is forced to be extremely small due to the constraint from the isocurvature perturbation, 
so a relatively large non-Gaussianity is expected.
In the case of $|\Psi(k_0)| \theta_i < H_I/2\pi$, the non-linearity parameter in our model is calculated as~\cite{Kawasaki:2008sn}
\be
	f_\mathrm{NL}^{\rm (iso)} = \frac{5}{162 \mathcal{P}_\zeta^{1/2}} \bigg( \frac{\mathcal{P}_\mathcal{S}}{\mathcal{P}_\zeta} \bigg)^{3/2} [\ln(kL)]^{-1/2},
	\label{f_NL}
\ee
where $L$ is an infrared cutoff which is taken to be the present Hubble horizon scale 
and we set $[\ln(kL)]^{1/2} = 5$~\cite{Lyth:2007jh}. 
The $f_\mathrm{NL}^{\rm (iso)}$~-~$f_a$ plane is shown in Fig.~\ref{Fig7b}.
It should be noticed that the $f_{\rm NL}^{\rm (iso)}$ we presented here is the non-Gaussianity in the 
(uncorrelated) CDM isocurvature perturbation~\cite{Kawasaki:2008sn,Kawasaki:2008pa,Langlois:2008vk,Hikage:2008sk,Hikage:2012be}, whose properties are different from non-Gaussianity in the adiabatic perturbation.
According to recent analysis~\cite{Hikage:2012be}, the constraint reads $f_{\rm NL}^{S,SS}\equiv (162/5)f_{\rm NL}^{\rm (iso)}
< 140$ at the 2$\sigma$ level. Thus our model can lead to a relatively large non-Gaussianity which may be close to the current observational bound.

\section{Conclusions} \label{conc}

We have shown that the smooth hybrid inflation naturally takes place 
in a SUSY axion model in which the PQ fields are identified with a part 
of the inflaton sector.
In order to reproduce the WMAP observation of the density perturbation, 
the PQ symmetry breaking scale must be of order of $10^{15}\,\mathrm{GeV}$.
The spectral index can naturally be the WMAP best fit value within 
the minimal K\"ahler potential.
Because the PQ symmetry is already broken during inflation, 
topological defects such as the cosmic strings and domain walls are never formed 
in this model.
After the inflation, the Universe is reheated by the decay of the heavy fields 
into the ordinary particles.
We have followed the dynamics of light scalar fields, saxion, after inflation,
and found that the saxion starts to oscillate with large initial amplitude of 
order $\alpha_s M_P$.
Thus the saxion eventually dominates the Universe and the decay of the saxion 
produces huge amount of entropy, which dilutes the gravitinos and axinos produced 
during reheating. 
The observed baryon asymmetry which survives the dilution can be generated through the Affleck-Dine mechanism.
The saxion can also produce Wino DM nonthermally with correct dark matter abundance 
if the Wino is much lighter than the gravitino,
as is expected from the AMSB or pure-gravity mediation model.
A severe constraint is imposed in this model due to the axion isocurvature perturbation, 
since the PQ symmetry is broken during inflation.
It excludes the possibility of axion coherent oscillation 
as a dominant component of the current CDM.
A non-Gaussianity of the isocurvature type, $f_\mathrm{NL}^{\rm (iso)} \sim 0.1 - 1$, 
is predicted in our model, which may be detected by the future observation.

\section*{Acknowledgment}

This work is supported by Grant-in-Aid for Scientific research from
the Ministry of Education, Science, Sports, and Culture (MEXT), Japan,
No.\ 14102004 (M.K.), No.\ 21111006 (M.K. and K.N.), No.\ 22244030 (K.N.) and also 
by World Premier International Research Center
Initiative (WPI Initiative), MEXT, Japan. 
N.K. is supported by the Japan Society for the Promotion of Science (JSPS).


  

\end{document}